\documentclass[a4paper,11pt]{article}

\pdfoutput=1 

\usepackage{jheppub} 

\usepackage{slashed}
\usepackage{subcaption}
\usepackage{xcolor}
\usepackage{cleveref}
\usepackage[T1]{fontenc} 
\usepackage[utf8]{inputenc}
\usepackage{mathtools}
\usepackage{bm}

\makeatletter
\newsavebox{\@brx}
\newcommand{\llangle}[1][]{\savebox{\@brx}{\(\m@th{#1\langle}\)}%
  \mathopen{\copy\@brx\kern-0.5\wd\@brx\usebox{\@brx}}}
\newcommand{\rrangle}[1][]{\savebox{\@brx}{\(\m@th{#1\rangle}\)}%
  \mathclose{\copy\@brx\kern-0.5\wd\@brx\usebox{\@brx}}}
\makeatother

\DeclarePairedDelimiter\floor{\lfloor}{\rfloor}

\title{\boldmath Tidal effects for spinning particles
}

\author[a]{Rafael Aoude,}
\author[b]{Kays Haddad,}
\author[c]{and Andreas Helset}
\affiliation[a]{Centre for Cosmology, Particle Physics and Phenomenology (CP3),\\
Universit\'{e} catholique de Louvain, 1348 Louvain-la-Neuve, Belgium}
\affiliation[b]{Niels Bohr International Academy and Discovery Center,
Niels Bohr Institute, \\ University of Copenhagen, Blegdamsvej 17,
DK-2100 Copenhagen, Denmark}
\affiliation[c]{Walter Burke Institute for Theoretical Physics,
California Institute of Technology,\\ Pasadena, CA 91125, USA}
\emailAdd{rafael.aoude@uclouvain.be}
\emailAdd{kays.haddad@nbi.ku.dk}
\emailAdd{andreas.helset@caltech.edu}

\abstract{Expanding on the recent derivation of tidal actions
for scalar particles, we present here the
action for a tidally deformed spin-$1/2$ particle.
Focusing on operators containing two powers of the Weyl tensor,
we combine the Hilbert series with an on-shell amplitude basis
to construct the tidal action.
With the tidal action in hand, we compute the leading-post-Minkowskian tidal
contributions to the spin-1/2 -- spin-1/2 amplitude, arising at $\mathcal{O}(G^{2})$.
Our amplitudes provide evidence that the observed long range spin-universality
for the scattering of two point particles extends to the scattering of tidally deformed objects.
From the scattering amplitude we find the conservative two-body Hamiltonian, linear and angular impulses, eikonal phase, spin kick, and aligned-spin scattering angle.
We present analogous results in the electromagnetic case along the way.}

\subheader{\normalsize
        \vspace*{-3.7em}
        \begin{flushright}
	CP3-20-57\\
        SAGEX-20-26-E\\
	CALT-TH/2020-055
        \end{flushright}
}

\allowdisplaybreaks

\begin{document}
\maketitle
\flushbottom

\section{Introduction}\label{sec:Intro}

The surge of attention payed to the binary inspiral problem in general relativity (GR)
by the scattering amplitudes community -- caused by the observation of gravitational
waves by the LIGO/Virgo collaborations \cite{LIGOGW} -- has produced
a greater understanding of the description of classical properties of binary systems
using quantum-field-theoretic and amplitudes techniques.
To date, most of the work has focused on point particles. 
Some notable results in this direction include elucidating methods for converting
scattering amplitudes for relativistic point particles to classical observables 
\cite{Guevara:2017csg,Cheung:2018wkq,Kosower:2018adc,Cristofoli:2019neg,Maybee:2019jus,Bjerrum-Bohr:2019kec,Bern:2020buy,Mogull:2020sak},
the description of classical angular momentum from quantum mechanical spin \cite{Guevara:2017csg,Guevara:2018wpp,Chung:2018kqs,Maybee:2019jus,Guevara:2019fsj,Arkani-Hamed:2019ymq,Chung:2019duq,Damgaard:2019lfh,Aoude:2020onz,Chung:2020rrz,Bern:2020buy},
and the state-of-the-art computation of the third post-Minkowskian (3PM) dynamics of a spinless
binary system \cite{Zvi3PM,Bern:2019crd,Cheung:2020gyp}.
Despite tremendous progress, there remains much to be understood about the
connection between scattering amplitudes and classical binary systems.

The quantum description of tidal effects is a topic of particular interest
of late.
While Schwarzschild
black holes do not tidally deform in four spacetime dimensions \cite{Binnington:2009bb,Damour:2009vw,Kol:2011vg,Gurlebeck:2015xpa}, there is still debate whether
the same is true for Kerr black holes in a general gravitational environment \cite{Pani:2015hfa,Landry:2015zfa,LeTiec:2020spy,LeTiec:2020bos}.
Nevertheless, such effects impact the gravitational wave signal of a neutron
star merger \cite{PhysRevLett.119.161101,PhysRevLett.121.161101},
so understanding them is necessary for the full description of these systems. 
The study of such effects from a scattering amplitudes perspective
was only recently initiated in ref.~\cite{Cheung:2020sdj}.
There, the authors included higher-dimensional operators quadratic in the Weyl tensor
in the action of a gravitating scalar particle, and computed the corrections to
the Hamiltonian and scattering angle up to next-to-leading post-Minkowskian (PM) order ($\mathcal{O}(G^{3})$ for tidal effects).
Not long after, two of the present authors applied the Hilbert series to extend the action of ref.~\cite{Cheung:2020sdj}
to include a complete, non-redundant set of operators quadratic in the Weyl tensor \cite{Haddad:2020que}.
They subsequently computed all $\mathcal{O}(G^2)$ finite-size\footnote{We will use the terms "tidal" and "finite-size" interchangably for both gravity and electromagnetism, even though "electric/magnetic susceptibility" is normally used in the electromagnetic context.} contributions to scalar-scalar scattering, including all contributions from higher-derivative operators.
Since then, refs.~\cite{Cheung:2020gbf,Bern:2020uwk} extended the study of finite-size effects by calculating the leading-PM contributions from an infinite set of operators with higher powers of the Weyl tensor, using the geodesic equation and unitarity cuts, respectively.

There have also been efforts to derive fully relativistic information
about tidally deformed systems from purely classical frameworks.
Refs.~\cite{Bini:2020flp,Kalin:2020mvi} incorporated tidal effects
into PM worldline actions, and subsequently derived the leading-PM
contributions to observables from a subset of tidal effects
arising from couplings quadratic in the Weyl tensor.
Ref.~\cite{Kalin:2020lmz} extended these results to a larger subset of these tidal operators,
as well as to the next-to-leading PM order.
Most recently, ref.~\cite{Gupta:2020lnv} presented a relativistic action
describing tidally deformed bodies up to linear order in spin.

To date, there has been no amplitudes approach
adding spin to the tidally deformed object.
In this paper we fill this gap by expanding on the work of ref.~\cite{Haddad:2020que} to include spin effects.
Combining the Hilbert series with on-shell methods, we construct the full
action for spinors and two powers of the Weyl tensor.
This allows us to compute all classical tidal effects at $\mathcal{O}(G^{2})$ for
spinor-spinor scattering.
Adapting the spinning effective field theory matching of ref.~\cite{Bern:2020buy},
we present the interaction Hamiltonian including spin and tidal effects at
$\mathcal{O}(G^{2})$ and to linear order in the angular momentum of each body.
Then, applying the methods of refs.~\cite{Kosower:2018adc,Maybee:2019jus}, we
use the amplitudes to find the linear and angular impulses.
We present the analogous results for quantum electrodynamics (QED).
We then compute the eikonal phase to extract additional observables in the gravitational case.
First, the eikonal phase allows us to verify the linear impulse through a separate method.
Then, it provides a means for computing the spin kick and the tidal
corrections to the aligned-spin scattering angle.

This paper is organized as follows.
We start by finding the tidal actions in both the electromagnetic and gravitational cases
for a massive spin-1/2 particle in \cref{sec:ampsNops}.
This is accomplished by first using the Hilbert series in \cref{sec:hilbertSeries}, which
provides a guide for finding the amplitude bases in \cref{sec:ampBasisQED,sec:ampBasisGR} and the operator bases in \cref{sec:opBasisQED,sec:opBasisGR}.
We then calculate the leading-PM scattering amplitudes in \cref{sec:leadingPM}.
The scattering amplitudes are then used to calculate various classical quantities:
the conservative Hamiltonian in \cref{sec:Hamiltonian},
linear and angular impulses in \cref{sec:impulse},  and the eikonal phase, spin kick, and aligned-spin scattering angle in \cref{sec:eikonal}. 
We conclude in \cref{sec:conclusion}. \Cref{sec:integral} contains a discussion on the
relevant loop integrals, while in \cref{app:fourierInt} we show details for the calculation of classical observables.

\section{Tidal actions}\label{sec:ampsNops}

Combining the Hilbert series with on-shell amplitudes methods, we construct
in this section the full action coupling two photon field strengths
or two Weyl tensors to spinor fields.

\subsection{Hilbert series}\label{sec:hilbertSeries}

We begin with the Hilbert series.
The Hilbert series produces the number of group invariants for a given field content, and
it is useful when constructing an operator basis in an effective field theory.
Notable achievements are the applications of the Hilbert series to the Standard Model effective field theory \cite{Lehman:2015via,Lehman:2015coa,Henning:2015alf,Henning:2015daa,Henning:2017fpj}, and the extension to include gravity \cite{Ruhdorfer:2019qmk}.
Non-relativistic effective field theories \cite{Kobach:2017xkw,Kobach:2018nmt} and effective field theories with non-linearly realized symmetries \cite{Graf:2020yxt} can also be constructed using Hilbert series techniques.

The Hilbert series was applied to characterize tidal effects for post-Minkowskian scattering in ref.~\cite{Haddad:2020que}.
In addition to the structures described in appendix A of ref.~\cite{Haddad:2020que},
we need the group characters for left- and right-handed Weyl spinors (respectively $\psi$ and $\psi^{\dagger}$) \cite{Ruhdorfer:2019qmk},
\begin{align}
    \chi_{[3/2,{(1/2,0)}]}(\mathcal{D};x,y)&=\mathcal{D}^{3/2}P(\mathcal{D};x,y)\left[\chi_{(1/2,0)}(x,y)-\mathcal{D}\chi_{(0,1/2)}(x,y)\right], \\
    \chi_{[3/2,{(0,1/2)}]}(\mathcal{D};x,y)&=\mathcal{D}^{3/2}P(\mathcal{D};x,y)\left[\chi_{(0,1/2)}(x,y)-\mathcal{D}\chi_{(1/2,0)}(x,y)\right].
\end{align}
Moreover, in both the electromagnetic and gravitational cases, we assume the
spinor fields are charged under a $U(1)$ gauge group.
  Thus we also need the gauge group characters $\chi_{U(1)}(\alpha)= \alpha^Q$ for a particle with charge $Q$ and the corresponding Haar measure:
\begin{align}
    \int d\mu_{U(1)}&=\frac{1}{2\pi i}\oint_{|\alpha|=1}\frac{d\alpha}{\alpha}.
\end{align}
All other relevant information is given in ref.~\cite{Haddad:2020que}.

We can now compute the Hilbert series in which we are interested.
The Hilbert series for two field strengths coupled to spinors for mass dimension $d$, $\mathcal{H}_d^{F^2}$, is
\begin{align}
	\mathcal{H}_{7+2n}^{F^2} &=
	\floor{n/2+1} (F_L^2 + F_R^2)(\psi \psi^c + \psi^{c\dagger}\psi^\dagger)D^{2n}
	+ n F_L F_R (\psi \psi^c + \psi^{c\dagger} \psi^\dagger)D^{2n}\label{eq:OddPhotHS}
	\nonumber \\&
	+ \frac{1}{2}(1-(-1)^n)( F_L^2 \psi \psi^c + F_R^2 \psi^{c\dagger}\psi^\dagger)D^{2n},
	\\
	\mathcal{H}_{6+2n}^{F^2} &=
	\floor{n/2} (F_L^2 + F_R^2)( \psi \psi^\dagger + \psi^c \psi^{c\dagger})D^{2n-1}
	+ n F_L F_R (\psi \psi^\dagger + \psi^c \psi^{c\dagger})D^{2n-1}.\label{eq:EvenPhotHS}
\end{align}
In \cref{eq:OddPhotHS} we have $n\geq0$, whereas $n\geq1$ in \cref{eq:EvenPhotHS}.
Coupling two Weyl tensors to spinors, the Hilbert series for mass dimension $d$, $\mathcal{H}_d^{C^2}$, is
\begin{align}
	\mathcal{H}_{7+2n}^{C^2} &=
	\floor{n/2+1} (C_L^2 + C_R^2)(\psi \psi^c + \psi^{c\dagger}\psi^\dagger)D^{2n}
	+ (n-1) C_L C_R (\psi \psi^c + \psi^{c\dagger} \psi^\dagger)D^{2n} \label{eq:OddGRHS}
	\nonumber \\&
	+ \frac{1}{2}(1-(-1)^n)( C_L^2 \psi \psi^c + C_R^2 \psi^{c\dagger}\psi^\dagger)D^{2n},
	\\
	\mathcal{H}_{6+2n}^{C^2} &=
	\floor{n/2} (C_L^2 + C_R^2)( \psi \psi^\dagger + \psi^c \psi^{c\dagger})D^{2n-1}
	+ (n-1) C_L C_R (\psi \psi^\dagger + \psi^c \psi^{c\dagger})D^{2n-1}.\label{eq:EvenGRHS}
\end{align}
Once again, $n\geq0$ in \cref{eq:OddGRHS} and $n\geq1$ in \cref{eq:EvenGRHS}.

\Cref{eq:EvenPhotHS,eq:EvenGRHS} are the Hilbert series for even mass dimensions. These operators
do not have any analogs in the complex scalar case (which is a slight generalization of the real scalar case discussed in ref.~\cite{Haddad:2020que}), and we will see that they all contribute 
spin effects in the PM amplitudes.

The Hilbert series for odd mass dimensions in \cref{eq:OddPhotHS,eq:OddGRHS} are very similar to
the corresponding Hilbert series for complex scalars coupled to photons or gravitons, respectively
(up to a doubling of the number of terms coming from chiral fermions).
The main difference is the appearance of the additional term
\begin{align}
	\label{eq:HScurious}
	\frac{1}{2}(1-(-1)^n) ( F_L^2 \psi \psi^c + F_R^2 \psi^{c\dagger} \psi^\dagger)D^{2n},
\end{align}
or its analog for gravitons.
These terms are present for $d=9,13,17,\dots$ -- where they contribute spin effects -- but are absent for $d=11,15,19,\dots$
In the next section, we will see that this curious behavior can be understood using on-shell
spinor-helicity variables.

\subsection{Amplitude basis for QED}\label{sec:ampBasisQED}

A complementary approach to the characterization of tidal effects 
is the construction of on-shell amplitudes.
It will be useful to us as it will elucidate relations among operators
that are not obvious in an off-shell language.
The massive spinor-helicity formalism of ref.~\cite{Arkani-Hamed:2017jhn} is ideal
for our purposes, and we will make use of it to construct the on-shell amplitude basis.
The massive spinors are indicated by a bolding of the momentum labels, which
also represents a symmetrization over the massive particle's little group indices. 
See refs.~\cite{Durieux:2019siw,Durieux:2020gip,Li:2020gnx,Falkowski:2019zdo} for recent work constructing
on-shell amplitudes.

Our approach makes use of the spinor structures presented in ref.~\cite{Durieux:2019siw}.
To extend these results to higher mass dimensions,
the various spinor structures are multiplied by combinations of Mandelstam variables,
$s_{ij}\equiv(p_i + p_j)^2$, in a way that respects the Bose/Fermi statistics of the system.
At four-points there are two independent Mandelstam variables.
Labeling the two bosons as $1$ and $2$ and the two fermions as $3$ and $4$,
we work with the two combinations of Mandelstam variables $x=s_{12}$ and $y= s_{13}-s_{23}+s_{24}-s_{14}$.\footnote{In massive four-point amplitudes, we must include the mass of the fermions as a further independent structure. However, the mass can always be absorbed into a Wilson coefficient, changing the dimensionality of the amplitude under consideration. Therefore, at a fixed mass dimension, it is sufficient to construct the helicity amplitudes using only $x$ and $y$.}
These combinations manifest symmetry/antisymmetry under the separate exchanges $1\leftrightarrow 2$ and $3\leftrightarrow 4$.

We generate all higher-dimensional helicity amplitudes by multiplying the various spinor structures by products of Mandelstam variables, e.g. $x^a y^b$ for $a,\ b$ non-negative integers.
As we are interested in amplitudes for two bosons and two spinors,
all amplitude structures must be symmetric (antisymmetric) under the exchange $1\leftrightarrow2$ ($3\leftrightarrow4$) for amplitudes with indistinguishable particles.
While Bose symmetry allows the power of $x$ in a helicity amplitude to be arbitrary, 
it restricts the power of $y$ in certain amplitudes to be either even or odd, i.e. some spinor structures will be multiplied by $x^a y^{2b(+1)}$.
Finally, ref.~\cite{Durieux:2020gip} argued that spinor structures for
massless particles can be generalized to the massive case by
simply bolding the momentum labels of the massive spin-1/2 particles.
All things considered, the amplitude basis for two massive spinors coupled to two photons is given in \cref{tab:ampBasisEM}.
\begin{table}[h!]
  \begin{center}
    \begin{tabular}{l|c}
      \textbf{Helicity} & \textbf{Amplitude} \\
      \hline
      $(++++)$ & $[12]^2 [\bm{34}] x^a y^{2b}$, \qquad $[12]([1\bm{4}][2\bm{3}] + [1\bm{3}][2\bm{4}]) y^{2b+1}$  \\
      $(++--)$ & $[12]^2\langle \bm{34}\rangle x^{a}y^{2b} $ \\
      $(+-++)$ & $[1|(\bm{3}-\bm{4})|2\rangle^2 [\bm{34}] x^a y^b$  \\
    $(+++-)$ & $[12]^2 [\bm{3}|(1-2)|\bm{4}\rangle x^a y^{2b+1}$  \\
      $(++-+)$ & $[12]^2 \langle\bm{3}|(1-2)|\bm{4}] x^a y^{2b+1}$  \\
      $(+-+-)$ & $[1\bm{3}]\langle 2\bm{4}\rangle [1|(\bm{3}-\bm{4})|2\rangle x^a y^b$ \\
    \end{tabular}
    \caption{The amplitude basis for electromagnetic finite-size effects. The helicity labels are ordered as $(\gamma_{1}\gamma_{2}\psi_{3}\psi_{4})$. The three first rows are the amplitude basis for odd mass dimensions, while the three last rows are for even mass dimensions. Here $a$ and $b$ take integer values from $0$ to $\infty$. The amplitudes for opposite helicities can be obtained by exchanging angle and square brackets.}
    \label{tab:ampBasisEM}
  \end{center}
\end{table}

We are now in a position to discuss the curious operators counted in \cref{eq:HScurious}.
They correspond to the second helicity amplitude in the first row of \cref{tab:ampBasisEM}.
The reason why they are only present for $d=9,13,17,\dots$ is a special relation between the
helicity amplitudes.
The helicity amplitude structures (considering massless fermions for simplicity)
\begin{align}
	\label{eq:independentHelicityAmplitudes}
	&[12]^2 [34] x^a y^{2b} \qquad\qquad \textrm{and}\qquad\qquad
	[12]([14][23]+[13][24]) y^{2c+1}
\end{align}
are independent for any $a,b,c$.
However, if we multiply the second helicity amplitude in \cref{eq:independentHelicityAmplitudes} by $x$, then we obtain the relation
\begin{align}\label{eq:OnShellStructureRelation}
	2[12]([14][23]+[13][24]) x y^{2c+1} = - [12]^2 [34] y^{2c+2}.
\end{align}
In the massive case this equivalence is modified by a lower-dimensional spinor structure.
Thus, we can choose an amplitude basis where the second term in \cref{eq:independentHelicityAmplitudes} is never multiplied by $x$.
Note that the analogous relation holds for gravitons, where each helicity amplitude is multiplied by $[12]^2$. The relation remains true when exchanging the square brackets for angle brackets.

\subsection{Amplitude basis for gravity}\label{sec:ampBasisGR}
The amplitude basis for gravity is almost identical to the photon case, with some additional powers
of $[12]$, $\langle1|(\bm{3}-\bm{4})|2]$, or their conjugates, 
accounting for the additional little group weights of gravitons relative to photons.
The full amplitude basis for gravity is listed in \cref{tab:ampBasisGR}.
\begin{table}[h!]
  \begin{center}
    \begin{tabular}{l|c}
      \textbf{Helicity} & \textbf{Amplitude} \\
      \hline
      $(++++)$ & $[12]^4 [\bm{34}] x^a y^{2b}$, \qquad $[12]^3([1\bm{4}][2\bm{3}] + [1\bm{3}][2\bm{4}]) y^{2b+1}$  \\
      $(++--)$ & $[12]^4\langle \bm{34}\rangle x^{a}y^{2b} $ \\
      $(+-++)$ & $[1|(\bm{3}-\bm{4})|2\rangle^4 [\bm{34}] x^a y^b$  \\
    $(+++-)$ & $[12]^4 [\bm{3}|(1-2)|\bm{4}\rangle x^a y^{2b+1}$  \\
      $(++-+)$ & $[12]^4 \langle\bm{3}|(1-2)|\bm{4}] x^a y^{2b+1}$  \\
      $(+-+-)$ & $[1\bm{3}]\langle 2\bm{4}\rangle [1|(\bm{3}-\bm{4})|2\rangle^3 x^a y^b$ \\
    \end{tabular}
    \caption{The amplitude basis for gravitational tidal effects. The helicity labels are ordered as $(g_{1}g_{2}\psi_{3}\psi_{4})$. The three first rows are the amplitude basis for odd mass dimensions, while the three last rows are for even mass dimensions. Here $a$ and $b$ take integer values from $0$ to $\infty$. The amplitudes for opposite helicities can be obtained by exchanging angle and square brackets.}
    \label{tab:ampBasisGR}
  \end{center}
\end{table}

\subsection{Operator basis for QED}\label{sec:opBasisQED}

With the explicit amplitude basis at hand, we can turn to finding the corresponding operator basis.
It can be beneficial to have both an amplitude and an operator basis, since then both on-shell and off-shell calculations can be performed directly starting from the appropriate basis.
Either approach can be used to calculate leading or subleading PM amplitudes.

In our case, the amplitude basis serves as a guide and as a cross-check. The products of Mandelstam variables
correspond to the distribution of covariant derivatives in the operators, and the
spinor structure can be simply found by putting various operators on-shell.
Moreover, the relation in \cref{eq:OnShellStructureRelation} indicates a
relation between off-shell operators that we must take into account.
As a cross-check, we have verified that the operator basis below matches the amplitude basis in \cref{tab:ampBasisEM} when put on-shell.

The full Lagrangian for fermions coupled to two field strengths is
\begin{align}
	\mathcal{L}_{\rm QED} = \bar \psi( i\slashed D - m ) \psi + \Delta\mathcal{L}_{\rm QED}^{\rm odd} + \Delta\mathcal{L}_{\rm QED}^{\rm even},
\end{align}
where $\Delta\mathcal{L}_{\rm QED}^{\rm odd/even}$ are the contributions from
higher-dimensional operators at odd or even mass dimensions, respectively.
Throughout this paper, we will use the prefix $\Delta$ to denote tidal contributions,
unless otherwise stated.
The contribution to the Lagrangian at odd mass dimensions is
\begin{align}
	\label{eq:opEModd}
	\Delta \mathcal{L}_{\rm QED}^{\rm odd}
	&= \sum_{n=0}^{\infty}\sum_{k=0}^{\floor{n/2}}
	a_1^{(n,k)} \left(\bar\psi \overset{\leftrightarrow}{D}\,^{\alpha_1\dots\alpha_{2k}}\psi \right) \left(D_{\beta_1\dots\beta_{n-2k}}F^{\mu\nu}\overset{\leftrightarrow}{D}_{\alpha_1\dots\alpha_{2k} }D^{\beta_1\dots\beta_{n-2k}}F_{\mu\nu} \right) \nonumber \\
	&+ \sum_{n=0}^{\infty}\sum_{k=0}^{\floor{n/2}}
	a_2^{(n,k)} \left(\bar\psi \overset{\leftrightarrow}{D}\,^{\mu\nu\alpha_1\dots\alpha_{2k}}\psi \right) \left(D_{\beta_1\dots\beta_{n-2k}}F_{\mu}^{\,\,\,\rho}\overset{\leftrightarrow}{D}_{\alpha_1\dots\alpha_{2k} }D^{\beta_1\dots\beta_{n-2k}}F_{\nu\rho} \right) \nonumber \\
	&+ \sum_{n=0}^{\infty}\sum_{k=0}^{\floor{n/2}}
	ia_3^{(n,k)} \left(\bar\psi \gamma_5 \overset{\leftrightarrow}{D}\,^{\mu\nu\alpha_1\dots\alpha_{2k+1}}\psi \right) \left(D_{\beta_1\dots\beta_{n-2k}}F_{\mu}^{\,\,\,\rho}\overset{\leftrightarrow}{D}_{\alpha_1\dots\alpha_{2k+1} }D^{\beta_1\dots\beta_{n-2k}}\tilde F_{\nu\rho} \right) \nonumber \\
	&+ \sum_{n=0}^{\infty}\sum_{k=0}^{\floor{n/2}}
	ia_4^{(n,k)} \left(\bar\psi \gamma_5\overset{\leftrightarrow}{D}\,^{\alpha_1\dots\alpha_{2k}}\psi \right) \left(D_{\beta_1\dots\beta_{n-2k}}F^{\mu\nu}\overset{\leftrightarrow}{D}_{\alpha_1\dots\alpha_{2k} }D^{\beta_1\dots\beta_{n-2k}}\tilde F_{\mu\nu} \right) \nonumber \\
	&+ \sum_{n=0}^{\infty}
i b^{(n)} \left(\bar\psi \sigma^{\mu\nu}\overset{\leftrightarrow}{D}\,^{\rho\alpha_1\dots\alpha_{2n}}\psi \right) \left(F_{\mu\rho}\overset{\leftrightarrow}{D}_{\sigma\alpha_1\dots\alpha_{2n} }F_{\nu}\,^{\sigma} \right)  .
\end{align}
We have only included parity-even operators. We have used the short-hand notation
$D^{\mu_1\dots\mu_k}=D^{\mu_1}\dots D^{\mu_k}$ and
$A \overset{\leftrightarrow}{D}\,^\mu B = A (D^\mu B) - (D^\mu A) B$.
In particular, $A \overset{\leftrightarrow}{D}\,^{\mu_1\dots\mu_k} B = A \overset{\leftrightarrow}{D}\,^{\mu_1\dots\mu_{k-1}}(D^{\mu_k} B) - (D^{\mu_k} A) \overset{\leftrightarrow}{D}\,^{\mu_1\dots\mu_{k-1}} B + \mathcal{O}(F^3)$. For our purposes, we only need the part quadratic in the field strengths or Weyl tensors.

The operators labelled by the Wilson coefficient $b^{(n)}$ produce the on-shell
structure on the right of \cref{eq:independentHelicityAmplitudes}.
As a consequence of \cref{eq:OnShellStructureRelation},
these operators only arise at every second odd mass dimension.

The contribution at even mass dimensions is
\begin{align}
	\label{eq:opEMeven}
	\Delta \mathcal{L}_{\rm QED}^{\rm even} 
	&= \sum_{n=0}^{\infty}\sum_{k=0}^{\floor{n/2}}
	i c_1^{(n,k)} \left(\bar\psi \gamma_\mu \overset{\leftrightarrow}{D}\,^{\nu\alpha_1\dots\alpha_{2k}}\psi \right) \left(D_{\beta_1\dots\beta_{n-2k}}F^{\mu\rho} \overset{\leftrightarrow}{D}_{\alpha_1\dots\alpha_{2k} }D^{\beta_1\dots\beta_{n-2k}}F_{\nu\rho} \right) \nonumber \\
	&+ \sum_{n=0}^{\infty}\sum_{k=0}^{\floor{n/2}}
	i c_2^{(n,k)} \left(\bar\psi \gamma_\mu \overset{\leftrightarrow}{D}\,^{\nu\lambda\alpha_1\dots\alpha_{2k}}\psi \right) \left(D_{\beta_1\dots\beta_{n-2k}}F^{\mu\rho} \overset{\leftrightarrow}{D}_{\lambda\alpha_1\dots\alpha_{2k} }D^{\beta_1\dots\beta_{n-2k}}F_{\nu\rho} \right) \nonumber \\
	&+ \sum_{n=0}^{\infty}\sum_{k=0}^{\floor{n/2}}
	c_3^{(n,k)} \left(\bar\psi \gamma_5\gamma_\mu \overset{\leftrightarrow}{D}\,^{\nu\lambda\alpha_1\dots\alpha_{2k}}\psi \right) \left(D_{\beta_1\dots\beta_{n-2k}}F^{\mu\rho} \overset{\leftrightarrow}{D}_{\lambda\alpha_1\dots\alpha_{2k} }D^{\beta_1\dots\beta_{n-2k}}\tilde F_{\nu\rho} \right. \nonumber \\ &\qquad\qquad\qquad\qquad\qquad\qquad\qquad\qquad
\left. - D_{\beta_1\dots\beta_{n-2k}}\tilde F^{\mu\rho} \overset{\leftrightarrow}{D}_{\lambda\alpha_1\dots\alpha_{2k} }D^{\beta_1\dots\beta_{n-2k}}F_{\nu\rho} \right) \nonumber \\
	&+ \sum_{n=0}^{\infty}\sum_{k=0}^{\floor{n/2}}
	c_4^{(n,k)} \left(\bar\psi \gamma_5\gamma_\mu \overset{\leftrightarrow}{D}\,^{\nu\lambda\alpha_1\dots\alpha_{2k}}\psi \right) \left(D_{\beta_1\dots\beta_{n-2k}}F^{\mu\rho} \overset{\leftrightarrow}{D}_{\lambda\alpha_1\dots\alpha_{2k} }D^{\beta_1\dots\beta_{n-2k}}\tilde F_{\nu\rho} \right. \nonumber \\ &\qquad\qquad\qquad\qquad\qquad\qquad\qquad\qquad
\left. + D_{\beta_1\dots\beta_{n-2k}}\tilde F^{\mu\rho} \overset{\leftrightarrow}{D}_{\lambda\alpha_1\dots\alpha_{2k} }D^{\beta_1\dots\beta_{n-2k}}F_{\nu\rho} \right)  ,
\end{align}
where again we only list the parity-even operators.

\subsection{Operator basis for gravity}\label{sec:opBasisGR}

We turn now to the tidal action for gravity.
The operator basis for fermions coupled to gravitons is very similar to the electromagnetic case.
We have verified that our operator basis produces the helicity amplitudes
in \cref{tab:ampBasisGR} when placed on-shell.

The full gravitational action includes the minimal coupling for fermions as well as the tidal perturbations to be described:
\begin{align}\label{eq:GRAction}
	\sqrt{-g} \mathcal{L}_{\rm GR} = \sqrt{-g}\left[\bar \psi( ie^\mu_{\,\,\,a} \gamma^a D_\mu - m)\psi
		+ \Delta\mathcal{L}_{\rm GR}^{\rm odd}
	+ \Delta\mathcal{L}_{\rm GR}^{\rm even} \right],
\end{align}
where the first part of the action is described in detail in e.g. ref.~\cite{Damgaard:2019lfh}.
The tidal contribution at odd mass dimensions is
\begin{align}
	\label{eq:opGRodd}
	\Delta \mathcal{L}_{\rm GR}^{\rm odd} 
	&= \sum_{n=0}^{\infty}\sum_{k=0}^{\floor{n/2}}
	d_1^{(n,k)} \left(\bar\psi \overset{\leftrightarrow}{D}\,^{\alpha_1\dots\alpha_{2k}}\psi \right) \left(D_{\beta_1\dots\beta_{n-2k}}C^{\mu\nu\rho\sigma}\overset{\leftrightarrow}{D}_{\alpha_1\dots\alpha_{2k} }D^{\beta_1\dots\beta_{n-2k}}C_{\mu\nu\rho\sigma} \right) \nonumber \\
	&+ \sum_{n=0}^{\infty}\sum_{k=0}^{\floor{n/2}}
	d_2^{(n,k)} \left(\bar\psi \overset{\leftrightarrow}{D}\,^{\mu\nu\lambda\tau\alpha_1\dots\alpha_{2k}}\psi \right) \left(D_{\beta_1\dots\beta_{n-2k}}C_{\mu\rho\lambda\sigma}\overset{\leftrightarrow}{D}_{\alpha_1\dots\alpha_{2k} }D^{\beta_1\dots\beta_{n-2k}}C_{\nu\rho\tau\sigma} \right) \nonumber \\
	&+ \sum_{n=0}^{\infty}\sum_{k=0}^{\floor{n/2}}
	id_3^{(n,k)} \left(\bar\psi \gamma_5\overset{\leftrightarrow}{D}\,^{\mu\nu\lambda\tau\alpha_1\dots\alpha_{2k+1}}\psi \right) \left(D_{\beta_1\dots\beta_{n-2k}}C_{\mu\rho\lambda\sigma}\overset{\leftrightarrow}{D}_{\alpha_1\dots\alpha_{2k+1} }D^{\beta_1\dots\beta_{n-2k}}\tilde C_{\nu\rho\tau\sigma} \right) \nonumber \\
	&+ \sum_{n=0}^{\infty}\sum_{k=0}^{\floor{n/2}}
	id_4^{(n,k)} \left(\bar\psi \gamma_5\overset{\leftrightarrow}{D}\,^{\alpha_1\dots\alpha_{2k}}\psi \right) \left(D_{\beta_1\dots\beta_{n-2k}}C^{\mu\nu\rho\sigma}\overset{\leftrightarrow}{D}_{\alpha_1\dots\alpha_{2k} }D^{\beta_1\dots\beta_{n-2k}}\tilde C_{\mu\nu\rho\sigma} \right) \nonumber \\
	&+ \sum_{n=0}^{\infty}
ie^{(n)} \left(\bar\psi \sigma^{\mu\nu}\overset{\leftrightarrow}{D}\,^{\rho\alpha_1\dots\alpha_{2n}}\psi \right) \left(C_{\mu\lambda\rho\tau}\overset{\leftrightarrow}{D}_{\sigma\alpha_1\dots\alpha_{2n} }C_{\nu}\,^{\lambda\sigma\tau} \right)  .
\end{align}
The operators labelled by the Wilson coefficient $e^{(n)}$ produce the gravitational analog of the on-shell
structure on the right of \cref{eq:independentHelicityAmplitudes}.
Again, \cref{eq:OnShellStructureRelation}  means that
these operators only arise at every second odd mass dimension.

The tidal contribution at even mass dimensions is
\begin{align}
	\label{eq:opGReven}
	\Delta \mathcal{L}_{\rm GR}^{\rm even} 
	&= \sum_{n=0}^{\infty}\sum_{k=0}^{\floor{n/2}}
	i f_1^{(n,k)} \left(\bar\psi \gamma_\mu \overset{\leftrightarrow}{D}\,^{\nu\gamma\delta\alpha_1\dots\alpha_{2k}}\psi \right) \left(D_{\beta_1\dots\beta_{n-2k}}C^{\mu\rho\gamma\tau} \overset{\leftrightarrow}{D}_{\alpha_1\dots\alpha_{2k} }D^{\beta_1\dots\beta_{n-2k}}C_{\nu\rho\delta\tau} \right) \nonumber \\
	&+ \sum_{n=0}^{\infty}\sum_{k=0}^{\floor{n/2}}
	i f_2^{(n,k)} \left(\bar\psi \gamma_\mu \overset{\leftrightarrow}{D}\,^{\nu\gamma\alpha_1\dots\alpha_{2k}}\psi \right) \left(D_{\beta_1\dots\beta_{n-2k}}C^{\mu\rho\gamma\tau} \overset{\leftrightarrow}{D}_{\delta\alpha_1\dots\alpha_{2k} }D^{\beta_1\dots\beta_{n-2k}}C_{\nu\rho\delta\tau} \right) \nonumber \\
	&+ \sum_{n=0}^{\infty}\sum_{k=0}^{\floor{n/2}}
	 f_3^{(n,k)} \left(\bar\psi \gamma_5\gamma_\mu \overset{\leftrightarrow}{D}\,^{\nu\gamma\delta\lambda\alpha_1\dots\alpha_{2k}}\psi \right) \left(D_{\beta_1\dots\beta_{n-2k}}C^{\mu\rho\gamma\tau} \overset{\leftrightarrow}{D}_{\lambda\alpha_1\dots\alpha_{2k} }D^{\beta_1\dots\beta_{n-2k}}\tilde C_{\nu\rho\delta\tau} \right. \nonumber \\ &\qquad\qquad\qquad\qquad\qquad\qquad\qquad\qquad
\left. - D_{\beta_1\dots\beta_{n-2k}}\tilde C^{\mu\rho\gamma\tau} \overset{\leftrightarrow}{D}_{\lambda\alpha_1\dots\alpha_{2k} }D^{\beta_1\dots\beta_{n-2k}}C_{\nu\rho\delta\tau} \right) \nonumber \\
	&+ \sum_{n=0}^{\infty}\sum_{k=0}^{\floor{n/2}}
	 f_4^{(n,k)} \left(\bar\psi \gamma_5\gamma_\mu \overset{\leftrightarrow}{D}\,^{\nu\lambda\alpha_1\dots\alpha_{2k}}\psi \right) \left(D_{\beta_1\dots\beta_{n-2k}}C^{\mu\rho\gamma\delta} \overset{\leftrightarrow}{D}_{\lambda\alpha_1\dots\alpha_{2k} }D^{\beta_1\dots\beta_{n-2k}}\tilde C_{\nu\rho\gamma\delta} \right. \nonumber \\ &\qquad\qquad\qquad\qquad\qquad\qquad\qquad\qquad
\left. + D_{\beta_1\dots\beta_{n-2k}}\tilde C^{\mu\rho\gamma\delta} \overset{\leftrightarrow}{D}_{\lambda\alpha_1\dots\alpha_{2k} }D^{\beta_1\dots\beta_{n-2k}}C_{\nu\rho\gamma\delta} \right) .
\end{align}
The form of the gravitational action is almost identical to the electromagnetic action. However, note that various
operators appear at different mass dimensions compared to the electromagnetic case, due to the additional Lorentz index
structure of the Weyl tensors.

\section{Leading-PM tidal effects}\label{sec:leadingPM}

\begin{figure}
    \centering
    \includegraphics[scale=1.2]{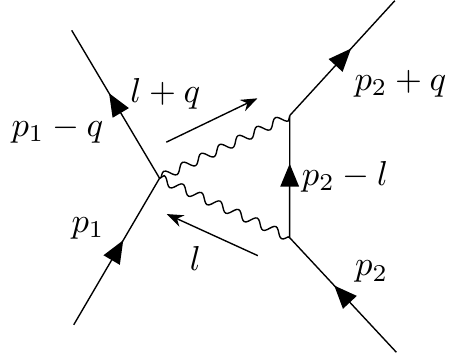}
    \caption{The only topology contributing classical tidal effects at one loop.
    Here, the tidal effects of particle 1 are probed.
    The wavy lines represent either photons or gravitons.}
    \label{fig:Triangle}
\end{figure}

The tidal
operators listed in \cref{eq:opEModd,eq:opEMeven,eq:opGRodd,eq:opGReven} are all we need
to compute the leading-PM tidal contributions to spin-1/2 -- spin-1/2 scattering.
There are no contributions at tree-level to the conservative $2\rightarrow2$ scattering amplitude, so
we must consider the scattering at one loop.
The only diagram contributing classically is the triangle diagram, shown in \cref{fig:Triangle},
where particle 1 is being tidally deformed.
Of course, the final result can be symmetrized in the particle labels to obtain the tidal deformation on particle 2.
We let the incoming momenta be $p_{i}^{\mu}=m_{i}v_{i}^{\mu}$, where $v_{i}^{\mu}$
are the particles' four-velocities, which satisfy $v_{i}^{2}=1$.
Also, we define $\omega\equiv v_{1}\cdot v_{2}$.
As we are interested in the classical portion of the amplitude, we compute
the leading-in-$\hbar$ contribution only.

We write the spin effects in terms of the covariant spin vector for heavy particles, defined as
\begin{align}
	\label{eq:spinDef}
    S_{i}^{\mu}&=\frac{1}{2}\bar{u}_{v_{i}}\gamma_{5}\gamma^{\mu}u_{v_{i}},
\end{align}
where $\gamma_{5}\equiv-i\gamma^{0}\gamma^{1}\gamma^{2}\gamma^{3}$ and $u_{v_{i}}$ are spinors for a heavy particle with velocity $v_{i}^{\mu}$ --- see refs.~\cite{Aoude:2020onz,Haddad:2020tvs} for their relation to Dirac spinors.
To do so, we convert all Dirac spinors to heavy spinors at the level of the on-shell amplitudes,
and keep only the terms at leading order in $\hbar$.
As argued in ref.~\cite{Aoude:2020onz}, this spin vector is identifiable with the
one-particle matrix element of the spin vector of a classical spinning object:
it automatically satisfies the covariant spin supplementary condition (SSC) $p_{i\mu}S^{\mu\nu}=0$ for a
spinning object with momentum $p_{i\mu}=m v_{i\mu}$, where $S^{\mu\nu}$ is the classical spin tensor \cite{tulczyjew1959motion}.\footnote{The momentum of a spinning object actually deviates from
$mv^{\mu}$ by corrections of $\mathcal{O}(RS^{2})$, where $R$ is a stand-in for the Riemann tensor \cite{Levi:2019kgk}.
This modifies the SSC at orders cubic in the object's angular momentum.
We can safely ignore such effects, as we are focused on contributions at most linear in the
spin of either object.}

The computation of all tidal effects requires 
knowledge of certain projections of the general-rank triangle integral.
Details about these integrals are given in \cref{sec:integral}.

\subsection{QED}\label{sec:leadingPMQED}

Since we are including spin effects for the tidally deformed particle, as well as for particle 2, the scattering amplitude will be decomposed in terms of spinless, spin-orbit, and spin-spin contributions.
In total, the finite-size contributions at one loop to the QED amplitude for spin-1/2 -- spin-1/2 scattering are
\begin{align}
	\label{eq:EMampResult}
        \Delta \mathcal{A}^{s=1/2}_{2} =& \frac{e^2 S}{\pi^2} \sum_{j=0}^{\infty}
        \left(-\frac{q^2}{2}\right)^{j+1} \left[ 
        \mathcal{U}_1 \mathcal{U}_2 F^{(0)}_j
        - i\omega \mathcal{E}_1\mathcal{U}_2 F_{j}^{(1,1)}
+ i\omega \mathcal{U}_1\mathcal{E}_2 F_{j}^{(1,2)} \right.\\&
\left.+ \left(q\cdot S_1\right)\left(  q\cdot S_2 \right)F_j^{(2,1)}
	-q^2\left(  S_1\cdot S_2\right)F_j^{(2,2)}
        + \omega q^2\left(v_2\cdot S_1\right)\left(  v_1\cdot S_2\right)F_j^{(2,3)}
        \right] , \nonumber 
\end{align}
where $e$ is the electromagnetic coupling, $S\equiv\pi^{2}/\sqrt{-q^{2}}$, and $S_{i}^{\mu}$ is the spin vector of particle $i$ defined in \cref{eq:spinDef}.
The subscript $2$ indicates that this is the amplitude at quadratic order in the coupling.
There are further spin structures that can appear at quadratic
order in spin, but they are subleading in the $\hbar$ expansion.
The form factors are functions of $\omega$, the dependence on which we leave implicit.
We find the form factors to be
\begin{align}
	F^{(0)}_j =& \sum_{k=0}^{j} a_1^{(j+k,k)} (4m_1)^{2k}(1-\omega^2)^{k}\frac{\alpha_k}{4}\nonumber\\
        +& \sum_{k=0}^{j} \left(a_2^{(j+k,k)}- \frac{c_1^{(j+k,k)}}{2m_1}\right) (4m_1)^{2k+2}(1-\omega^2)^{k}\frac{1}{32}\left[ (\omega^2-1)\alpha_{k+1} - \omega^2 \alpha_k  \right] , \\
	F_j^{(1,1)} =& 
 - \frac{b^{(j)}}{4m_2} (4m_1)^{2j}  (1-\omega^2)^j  \alpha_j \nonumber \\
	+&\sum_{k=0}^{j} \frac{c_1^{(j+k,k)} }{4m_1 m_2} \left(4 m_1\right)^{2k}  (1-\omega^2)^{k} \left[  \alpha_{k+1} - \alpha_k \right] \nonumber \\
        +&\sum_{k=0}^{j} \frac{c_4^{(j+k,k)}-c_3^{(j+k,k)}}{8m_1 m_2}  \left(4 m_1\right)^{2k+2} (1-\omega^2)^k \alpha_{k+1}  ,
	\\
	F_j^{(1,2)} =&  \sum_{k=0}^{j} \left(a_2^{(j+k,k)}- \frac{c_1^{(j+k,k)}}{2m_1}\right)  \frac{\left(4 m_1\right)^{2k+2}}{32m_1 m_2^2} 
            (1-\omega^2)^k   
        \left[ 2 \alpha_{k+1} -\alpha_{k} \right] ,
\\
	F_j^{(2,1)} =& 
\sum_{k=0}^{j} 
\frac{a_4^{(j+k,k)}}{4m_1 m_2}
   (4m_1)^{2 k} \left(1-\omega^2\right)^k    \alpha_k
	\nonumber \\ 
	+&\frac{b^{(j)}}{16 m_2}    (4m_1)^{2j+1}(1-\omega^2)^{j}
\left[
        ( 3\omega^2-2)(\alpha_{j}+\alpha_{j+1})   
+   (2\omega^2-1) 
\frac{2}{2j+1}\alpha_{j+1} 
\right] \nonumber\\
	     +& \sum_{k=0}^{j} \frac{c_1^{(j+k,k)}}{8 m_2}   \left(4 m_1\right)^{2k} (1-\omega^2)^k
\left[
               (2\omega^2-1) \left( \alpha_k - 2  \alpha_{k+1}     \right)
        \right]\nonumber \\ 
	     -&  \sum_{k=0}^{j} \frac{c_{3}^{(j+k,k)}}{16m_2} (4m_1)^{2k+2} (1-\omega^2)^{k+1 }
\frac{1}{k+2} \alpha_{k+1}  \nonumber \\
		    +&\sum_{k=0}^{j} \frac{c_4^{(j+k,k)}}{8m_2}  \left(4 m_1\right)^{2k+2}
         (1- \omega^2)^{k+1}  \alpha_{k+1} , \\
	F_j^{(2,2)} =& 
   \frac{b^{(j)}}{16 m_2}    (4m_1)^{2j+1}(1-\omega^2)^{j}
\left[
   (2\omega^2-1) 
\frac{1}{2j+1} \alpha_{j+1} 
\right] \nonumber \\
	     +&\sum_{k=0}^{j} \frac{c_1^{(j+k,k)}}{8 m_2}   \left(4 m_1\right)^{2k} (1-\omega^2)^k
\left[
               (2\omega^2-1) \left( \alpha_k - 2  \alpha_{k+1}     
                      \right)
        \right]\nonumber \\ 
	     -&  \sum_{k=0}^{j} \frac{c_{3}^{(j+k,k)}}{16m_2} (4m_1)^{2k+2} (1-\omega^2)^{k+1 }
\frac{1}{k+2} \alpha_{k+1} , \\
	F_j^{(2,3)} =& 
   \frac{b^{(j)}}{32 m_2}    (4m_1)^{2j+1}(1-\omega^2)^{j}
        \left(4\alpha_{j+1}   
-    (2\omega^2-1) 
\alpha_{j+1}\frac{4j}{2j+1}     \frac{1}{\omega^2-1} \right)
 \nonumber \\
+&   \sum_{k=0}^{j} \frac{c_1^{(j+k,k)}}{8 m_2}   \left(4 m_1\right)^{2k}  (1-\omega^2)^k
\left[
 \alpha_k 
        - \frac{4k}{2k+1}  \frac{\omega^2}{ (\omega^2-1)}\alpha_{k+1} 
        \right]\nonumber \\ 
		     +&\sum_{k=0}^{j} \frac{c_{3}^{(j+k,k)}}{4m_2} (4m_1)^{2k+2} (1-\omega^2)^k 
\left(\alpha_{k+1}  - \frac{2(k+1)}{2k+3}\alpha_{k+2}\right).
\end{align}
We have introduced the notation
\begin{align}
	\alpha_k \equiv \frac{\left(\frac{1}{2}\right)_k}{2^k (1)_k},
\end{align}
where $(a)_{k}$ is the Pochhammer symbol, and
\begin{align}
	u_{v_{1}} \equiv u(m_1 v_1), \quad
	u_{v_{2}} \equiv u(m_2 v_2), &\quad
	\bar u_{v_{1}} \equiv \bar u(m_1 v_1-q), \quad
	\bar u_{v_{2}} \equiv \bar u(m_2 v_2+q), \\
	\bar u_{v_{1}} u_{v_{1}} \equiv \mathcal{U}_1, &\quad
	\bar u_{v_{2}} u_{v_{2}} \equiv \mathcal{U}_2, \\
	v_{2\mu}q_{\nu}\bar{u}_{v_{1}}\sigma^{\mu\nu}u_{v_{1}} = -2&v_{2\mu}q_{\nu} \epsilon^{\mu\nu\alpha\beta}v_{1\alpha} S_{1\beta}
	 \equiv- 2 \mathcal{E}_1 /(m_1 m_2) \label{eq:Spin1Ident}, \\
	v_{1\mu}q_{\nu}\bar{u}_{v_{2}}\sigma^{\mu\nu}u_{v_{2}} = -2&v_{1\mu}q_{\nu} \epsilon^{\mu\nu\alpha\beta}v_{2\alpha} S_{2\beta} 
	\equiv 2 \mathcal{E}_2 /(m_1 m_2) \label{eq:Spin2Ident}, 
\end{align}
%
to simplify the expressions, where the Levi-Civita tensor is defined by $\epsilon^{0123}=1$.
We remark on the absence of contributions from the operators with Wilson coefficients
$a_{3}^{(j+k,k)}$ and $c_{2}^{(j+k,k)}$:
the contributions from these operators scale as $(q^{2})^{j+2}$, so we treat them
as subleading in the $\hbar$ expansion.

Up to a redefinition of Wilson coefficients, the form-factor 
for the spin-monopole $F^{(0)}_{j}$ agrees with the amplitude
in ref.~\cite{Haddad:2020que}.
The matching conditions, accounting for the normalization of the spinors, are
\begin{align}
    4^{2k}a_{1}^{(j+k,k)}&\rightarrow \frac{a_{k}^{(j+k)}}{m_{1}}, \\
    4^{2k+2}\left(a_{2}^{(j+k,k)}-\frac{c_{1}^{(j+k,k)}}{2m_{1}}\right)&\rightarrow-\frac{4}{m_{1}}b_{k}^{(j+k+1)},
\end{align}
where the Wilson coefficients on the right hand side are those in ref.~\cite{Haddad:2020que}.
This suggests that the spin-multipole universality for long range 
electromagnetic scattering \cite{Holstein:2008sw} extends to the case where tidal
deformations are accounted for.
This means that, at least at the classical level, the tidally-modified 
spin-monopole is the same regardless of the total spin of the object.
The agreement with the scalar amplitude also shows that the additional even-dimensional operators
do not contribute unique spinless structures.

\subsection{Gravity}\label{sec:leadingPMgravity}

Again, we split the gravitational scattering amplitude in terms of spinless, spin-orbit, and spin-spin contributions.
The leading-in-$\hbar$ tidal contributions at 2PM order are
\begin{align}
	\label{eq:GRampResult}
        \Delta \mathcal{M}^{s=1/2}_{2} =& G^2 m_2^2 S \sum_{j=0}^{\infty}
        \left(-\frac{q^2}{2}\right)^{j+2} \left[ 
        \mathcal{U}_1 \mathcal{U}_2 G^{(0)}_j
        - i\omega \mathcal{E}_1\mathcal{U}_2 G_{j}^{(1,1)}
+ i\omega \mathcal{U}_1\mathcal{E}_2 G_{j}^{(1,2)} \right.\\ &
\left.+ \left(q\cdot S_1\right)\left(  q\cdot S_2 \right)G_j^{(2,1)}
	-q^2\left(  S_1\cdot S_2\right)G_j^{(2,2)}
        + \omega q^2\left(v_2\cdot S_1\right)\left(  v_1\cdot S_2\right)G_j^{(2,3)}
        \right] , \nonumber 
\end{align}
where the form factors are the following functions of $\omega$:
\begin{align}
	G^{(0)}_j =& \sum_{k=0}^{j}  d_1^{(j+k,k)} \left(4 m_1\right)^{2k} 
         (1-\omega^2)^k 16 \alpha_k  \nonumber \\
        +&   \sum_{k=0}^{j}\left( d_2^{(j+k,k)} - \frac{f_1^{(j+k,k)}}{2m_1} \right) \left(4 m_1\right)^{2k+4}
        (1-\omega^2)^k
        \nonumber \\ & \times
        \frac{1}{8}\left[
                 (1-2\omega^2)^2 \alpha_k    
                + 2  (1-4\omega^2) (\omega^2-1) \alpha_{k+1}  
                + 2  (\omega^2-1)^{2} \alpha_{k+2} 
        \right] ,\\
	G_j^{(1,1)}=& 
	-  \frac{ e^{(j)}}{m_2}  (4m_1)^{2j}   
         (1-\omega^2)^j
        16\alpha_{j+1} \nonumber \\
	+& \sum_{k=0}^{j} \frac{f_1^{(j+k,k)}}{m_1 m_2}  \left(4 m_1\right)^{2k+2}
          (1-\omega^2 )^{k}
       \nonumber \\ &\times
        \left[2(\omega^2 -1)( \alpha_{k+2}- 2 \alpha_{k+1}) 
         + (2\omega^2-1)  
                2\alpha_{k}  
         + (1-4\omega^2)  
                 \alpha_{k+1}  
        \right]  \nonumber \\
+& \sum_{k=0}^{j} \frac{f_3^{(j+k,k)} }{2m_1 m_2} (4m_1)^{2k+4}  (1-\omega^2)^{k}
        \left[
                  (4\omega^2-3)\alpha_{k+1}-6 (\omega^2-1)\alpha_{k+2}
        \right] \nonumber \\
  +&     \sum_{k=0}^{j} \frac{f_4^{(j+k,k)} }{m_1 m_2} \left(4 m_1\right)^{2k+2}
         (1-\omega^2)^k 8\alpha_{k+1} , \\
	\label{eq:GRampSpin2}
        G_j^{(1,2)}
	=&  \sum_{k=0}^{j} \left(d_2^{(j+k,k)}  -\frac{f_1^{(j+k,k)} }{2 m_2}\right) \frac{\left(4 m_1\right)^{2k+4}}{2m_1 m_2^2}
         (1-\omega^2)^k
        \nonumber \\ & \times
        \left[
         (\omega^2-1) (2\alpha_{k+2} - \alpha_{k+1} )
         - \frac{1}{2}(2\omega^2-1) (2 \alpha_{k+1}-  \alpha_k)   
        \right]  , \\
        G_j^{(2,1)} =& 
\sum_{k=0}^{j} 
\frac{d_4^{(j+k,k)}}{m_1 m_2}
   (4m_1)^{2 k}  \left(1-\omega^2\right)^k   16 \alpha_k
	\nonumber \\
	-&\frac{e^{(j)}}{m_2} (4m_1)^{2j+1}   (1-\omega^2)^{j+1} 4\alpha_{j+1}
            \nonumber \\
-& \sum_{k=0}^{j} 
\frac{f_1^{(j+k,k)}}{m_2}
  (4m_1)^{2 k+2}  \left(1-\omega^2\right)^k 
   \frac{4(2k+7)\omega^4 - 3(2k+9)\omega^2+3}{8(k+1)(k+2)}\alpha_k
	\nonumber \\
+& \sum_{k=0}^{j} 
        \frac{f_3^{(j+k,k)}}{m_2}
        (4m_1)^{2 k+4}    \left(1-\omega ^2\right)^{k+1} 
    \frac{3(2k+7)\omega^2 - 3}{8(k+2)(k+3)}\alpha_{k+1}
	\nonumber \\
+&\sum_{k=0}^{j} \frac{f_4^{(j+k,k)}}{m_2}  \left(4 m_1\right)^{2k+2}
         (1- \omega^2)^{k+1} 8 \alpha_{k+1} , \\
        G_j^{(2,2)} =& 
- \sum_{k=0}^{j} 
\frac{f_1^{(j+k,k)}}{m_2}
  (4m_1)^{2 k+2}  \left(1-\omega^2\right)^k 
   \frac{4(2k+7)\omega^4 - 3(2k+9)\omega^2+3}{8(k+1)(k+2)}\alpha_k
	\nonumber \\
+& \sum_{k=0}^{j} 
        \frac{f_3^{(j+k,k)}}{m_2}
        (4m_1)^{2 k+4}    \left(1-\omega ^2\right)^{k+1} 
    \frac{3(2k+7)\omega^2 - 3}{8(k+2)(k+3)}\alpha_{k+1}
   , \\
        G_j^{(2,3)} =& 
	\sum_{k=0}^{j} \frac{f_1^{(j+k,k)}}{m_2}  \left(4 m_1\right)^{2k+2}
         (1-\omega^2)^{k}
        \nonumber \\ &\times
        \left[
                \frac{\alpha_k}{4(k+1)(k+2)(\omega^2-1)} (-2(2k+7)\omega^4 + (k(2k+11)+17)\omega^2 - 3(k+1))
        \right] \nonumber \\
+&  \sum_{k=0}^{j} 
\frac{f_3^{(j+k,k)}}{m_2}
 (4 m_1)^{2 k+4}   \left(1-\omega^2\right)^k 
 \frac{3(k+4)-(k+6)(2k+7)\omega^2}{4(k+2)(k+3)} \alpha_{k+1}.
\end{align}
The contributions from the $d_{3}^{(j+k,k)}$ and $f_{2}^{(j+k,k)}$ operators scale 
as $(q^{2})^{j+3}$, so we treat them as subleading in the $\hbar$ expansion.

Results for the amplitude describing the scattering of a tidally defomed, spinless object 
with a spinning point particle were recently presented in ref.~\cite{Bern:2020uwk}.
These are to be compared with the function $G_{j}^{(1,2)}$. After some algebraic
manipulations of \cref{eq:GRampSpin2} we find agreement with eq.~(3.68) in ref.~\cite{Bern:2020uwk} when  $k=j$.

As in the electromagnetic case, the form-factor 
for the spin-monopole $G^{(0)}_{j}$ agrees with the amplitude
in ref.~\cite{Haddad:2020que}; the two sets of Wilson coefficients can be matched through
\begin{align}
	\label{eq:MonopoleAmpMatch1}
    4^{2k}d_{1}^{(j+k)}&\rightarrow\frac{c_{k}^{(j+k)}}{m_{1}}, \\
	\label{eq:MonopoleAmpMatch2}
    4^{2k+4}\left(d_{2}^{(j+k,k)}-\frac{f_{1}^{(j+k)}}{2m_{1}}\right)&\rightarrow\frac{16}{m_{1}}d_{k}^{(j+k+2)},
\end{align}
where the coefficients on the right hand side are
those in ref.~\cite{Haddad:2020que}.
Under these replacements the form factor here is related to that in ref.~\cite{Haddad:2020que} through
$G^{(0)}_{j}\rightarrow g_{j}/m_{1}$.
The differing mass dimensions of the Wilson coefficients in each action
is because of the different mass dimensions of scalar versus spinor fields.
This matching extends the spin-multipole universality for long range 
gravitational scattering observed in ref.~\cite{Holstein:2008sx} to the tidally deformed setting.
More precisely, this provides evidence that the classical tidally-modified
spin-monopole is the same regardless of the total spin of the deformed object.

In both the electromagnetic and gravitational point-particle cases, 
it is well known that classical spin-spin effects of the form $q^{2}\,S_{1}\cdot S_{2}$ and $q\cdot S_{1}\,q\cdot S_{2}$
arise in the proportion $q\cdot S_{1}\,q\cdot S_{2}-q^{2}\,S_{1}\cdot S_{2}$ through one-loop order \cite{Holstein:2008sw,Holstein:2008sx,Damgaard:2019lfh,Bern:2020buy}.
We have shown here that this correlation between spin structures is broken for general values of the
Wilson coefficients when finite-size effects are included at the one-loop level.

In the interest of deriving a conservative Hamiltonian
in \Cref{sec:Hamiltonian}, we rewrite the amplitude in the center-of-mass 
kinematics of refs.~\cite{Lorce:2017isp,Bern:2020buy}.
These are, for all momenta outgoing
\begin{align}
    p_{1}^{\mu}=-(E_{1},\bm{p}),\quad p_{2}^{\mu}=-(E_{2},-\bm{p}),&\quad q^{\mu}=(0,\bm{q}),\quad \bm{p}\cdot\bm{q}=\frac{\bm{q}^{2}}{2},\notag \\
    S_{1}^{\mu}=\left(\frac{\bm{p}\cdot\bm{S}_{1}}{m_{1}},\bm{S}_{1}+\frac{\bm{p}\cdot \bm{S}_{1}}{(E_{1}+m_{1})m_{1}}\bm{p}\right),&\quad S_{2}^{\mu}=\left(-\frac{\bm{p}\cdot\bm{S}_{2}}{m_{2}},\bm{S}_{2}+\frac{\bm{p}\cdot \bm{S}_{2}}{(E_{2}+m_{2})m_{2}}\bm{p}\right).
\end{align}
Here, $\bm{S}_{i}$ is the spin vector in the rest frame of particle $i$.
It is also the spatial component of the canonical spin vector, which is the spin vector appearing in the canonical Hamiltonian \cite{Vines:2017hyw}.
We express the spin structures that arise using these kinematics:\footnote{We use the mostly negative metric signature.
Note that we find the opposite sign on the first term of $q\cdot S_{i}$ relative to ref.~\cite{Bern:2020buy}. This difference is immaterial, however, since these structures only arise in the combination $q\cdot S_{1}\, q\cdot S_{2}$, causing the signs to cancel, and since the second term in this inner product is subleading in $\hbar$.}
\begin{align}
    q\cdot S_{i}=-\bm{q}\cdot\bm{S}_{i}-\frac{\bm{q}^{2}\bm{p}\cdot \bm{S}_{i}}{2m_{i}(E_{i}+m_{i})},&\quad \mathcal{E}_{i}=E\, (\bm{p}\times\bm{q})\cdot\bm{S}_{i},\notag \\
    p_{2}\cdot S_{1}=-\frac{E}{m_{1}}\bm{p}\cdot \bm{S}_{1},&\quad p_{1}\cdot S_{2}=\frac{E}{m_{2}}\bm{p}\cdot \bm{S}_{2},\notag \\
    S_{1}\cdot S_{2}=-\bm{S}_{1}\cdot\bm{S}_{2}&-M_{12}\bm{p}\cdot \bm{S}_{1}\,\bm{p}\cdot\bm{S}_{2},
\end{align}
where
\begin{align}
    M_{12}\equiv\frac{(E+M)^{2}}{2(E_{1}+m_{1})(E_{2}+m_{2})m_{1}m_{2}},
\end{align}
$E=E_{1}+E_{2}$, and $M=m_{1}+m_{2}$.
Substituting these into \cref{eq:GRampResult} after non-relativistically normalizing the amplitude,
and keeping only the leading-in-$\hbar$ terms, we find
\begin{align}
	\label{eq:AmpCovCoeffsF}
    \frac{\Delta\mathcal{M}^{s=1/2}_{2}}{4E_{1}E_{2}}&=
   2G^{2}S
    \left[\mathcal{U}_{1}\mathcal{U}_{2}\Delta a_{{\rm cov},2}^{(0)}
	    +\Delta a_{{\rm cov},2}^{(1,1)} \,i(\bm{p}\times\bm{q})\cdot\bm{S}_{1}\,\mathcal{U}_{2}+\Delta a_{{\rm cov},2}^{(1,2)}\, \mathcal{U}_{1}\,i(\bm{p}\times\bm{q})\cdot\bm{S}_{2}
	    \right.\nonumber \\ &
    \left.+\Delta a_{{\rm cov},2}^{(2,1)}\,(\bm{q}\cdot \bm{S}_1\,\bm{q}\cdot \bm{S}_2)
	+\Delta a_{{\rm cov},2}^{(2,2)}\,(\bm{q}^2\bm{S}_1\cdot \bm{S}_2)
	 +\Delta a_{{\rm cov},2}^{(2,3)}\,\bm{q}^{2}\bm{p}\cdot\bm{S}_{1}\,\bm{p}\cdot \bm{S}_{2}
	\right],
\end{align}
where
\begin{align}
	\label{eq:AmpCovCoeffs0}
    \Delta a_{{\rm cov},2}^{(0)}&=m_{2}^{2}\sum_{j=0}^{\infty}\left(\frac{\bm{q}^2}{2} \right)^{j+2}\left[\frac{G^{(0)}_{j}}{8E_{1}E_{2}}\right], \\
	\label{eq:AmpCovCoeffs11}
    \Delta a_{{\rm cov},2}^{(1,1)}&=m_{2}^{2}\sum_{j=0}^{\infty}\left(\frac{\bm{q}^2}{2} \right)^{j+2}\left[-\frac{\omega E}{8E_{1}E_{2}}G_{j}^{(1,1)}\right], \\
	\label{eq:AmpCovCoeffs12}
    \Delta a_{{\rm cov},2}^{(1,2)}&=m_{2}^{2}\sum_{j=0}^{\infty}\left(\frac{\bm{q}^2}{2} \right)^{j+2}\left[\frac{\omega E }{8E_{1}E_{2}}G_{j}^{(1,2)}\right], \\
	\label{eq:AmpCovCoeffs21}
    \Delta a_{{\rm cov},2}^{(2,1)}&=m_{2}^{2}\sum_{j=0}^{\infty}\left(\frac{\bm{q}^2}{2} \right)^{j+2}\left[\frac{1}{8E_{1}E_{2}}G_{j}^{(2,1)}\right], \\
	\label{eq:AmpCovCoeffs22}
    \Delta a_{{\rm cov},2}^{(2,2)}&=m_{2}^{2}\sum_{j=0}^{\infty}\left(\frac{\bm{q}^2}{2} \right)^{j+2}\left[-\frac{1}{8E_{1}E_{2}}G_{j}^{(2,2)}\right], \\
	\label{eq:AmpCovCoeffs23}
    \Delta a_{{\rm cov},2}^{(2,3)}&=m_{2}^{2}\sum_{j=0}^{\infty}\left(\frac{\bm{q}^2}{2} \right)^{j+2}\left[ \frac{\omega E^2}{8E_{1}E_{2}m_1^2 m_2^2} G_j^{(2,3)} - \frac{M_{12}}{8E_{1}E_{2}} G_j^{(2,2)}\right].
\end{align}
We have borrowed the notation from ref.~\cite{Bern:2020buy}, where the subscript ${\rm cov}$ denotes that these are the coefficients
to the spin structures when the amplitude is written
in terms of the covariant spin vectors.
We define the notation
\begin{align}\label{eq:CovCoeffNotation}
    \Delta a_{{\rm cov},2}^{A}\equiv m_{2}^{2}\sum_{j=0}^{\infty}\left(\frac{\bm{q}^{2}}{2}\right)^{j+2}\Delta a_{{\rm cov},2,j}^{A}(\omega),
\end{align}
for easy reference later on.

\section{Conservative two-body Hamiltonian}
\label{sec:Hamiltonian}

In this section, we use the effective field theory (EFT) matching approach of ref.~\cite{Bern:2020buy} to derive the two-body spin-dependent conservative Hamiltonian.
Working with the spin-coherent states $|\bm{n}\rangle$ defined therein,
the two-body Hamiltonian is given by
\begin{align}
H(\bm{q},\bm{p}) = \sqrt{\bm{p}^{2} + m_1^2} + \sqrt{\bm{p}^2+m_2^2} + \langle \bm{n}_1 \bm{n}_2 |\left[ \hat{V} (\bm{k}',\bm{k}, \hat{\bm{S}}_a) +  \Delta \hat{V} (\bm{k}',\bm{k}, \hat{\bm{S}}_a)\right]| \bm{n}_1\bm{n}_2 \rangle ,
\end{align}
where $\Delta\hat{V}$ encodes the tidal contributions to the Hamiltonian for spinning particles.
Here, $\bm{k}$ is the incoming three-momentum, 
$\bm{k}^{\prime}$ is the outgoing three-momentum,
$\bm{q}\equiv\bm{k}-\bm{k}^{\prime}$ is the transferred three-momentum,
and $\bm{p}\equiv(\bm{k}'+\bm{k})/2$. Finally, $\hat{\bm{S}}_a$ is the rest-frame spin operator of particle $a$,
whose expectation value in the spin-coherent state of particle $a$ gives its rest-frame spin vector, $\bm{S}_{a}$.
The tidal potential can be expanded in the basis of spin operators analogously to the expansion
of $\hat{V}(\bm{k}',\bm{k},\hat{\bm{S}}_{a})$ in ref.~\cite{Bern:2020buy}:
\begin{align}\label{eq:TidalSpinExp}
\Delta \hat{V}(\bm{k}',\bm{k}, \hat{\bm{S}}_a) = \sum_A \Delta\hat{V}^A (\bm{k}',\bm{k})\, \hat{\mathbb{O}}^A.
\end{align}
$A$ labels the following classical spin structures:
\begin{align}\label{eq:SpinBasis}
\hat{\mathbb{O}}^{(0)} = \mathbb{I}, \qquad
\hat{\mathbb{O}}^{(1,1)} &= \bm{L}_q \cdot \hat{\bm{S}}_1, \qquad \hat{\mathbb{O}}^{(1,2)} = \bm{L}_q \cdot \hat{\bm{S}}_2,\notag \\
\hat{\mathbb{O}}^{(2,1)} = \bm{q} \cdot \hat{\bm{S}}_1 \bm{q} \cdot \hat{\bm{S}}_2, \qquad
\hat{\mathbb{O}}^{(2,2)} &= \bm{q}^2 \hat{\bm{S}}_1 \cdot \hat{\bm{S}}_2, \qquad
\hat{\mathbb{O}}^{(2,3)} = \bm{q}^2 \bm{p} \cdot \hat{\bm{S}}_1 \bm{p} \cdot \hat{\bm{S}}_2,
\end{align}
where $\bm{L}_q \equiv i (\bm{p} \times \bm{q})$.
The first index in the superscripts labels the number of spin vectors in the operator,
whereas the second labels the different structures with that many spin vectors.
There are two spin structures that are not included in this basis.
They are
\begin{align}
    \bm{q}\cdot\bm{p}\, \bm{q}\cdot\hat{\bm{S}}_{1}\bm{p}\cdot\hat{\bm{S}}_{2},\quad \bm{q}\cdot\bm{p}\, \bm{p}\cdot\hat{\bm{S}}_{1}\bm{q}\cdot\hat{\bm{S}}_{2}.
\end{align}
As discussed in ref.~\cite{Bern:2020buy}, these are omitted from 
the basis since the on-shell condition $\bm{q}\cdot\bm{p}\sim\bm{q}^{2}$
means they are subleading in the $\hbar$ expansion.

To match the tidal amplitude to the tidal potential,
the coefficients of these operators are also expanded in powers of $G$:
\begin{align}\label{eq:TidalPMExp}
	\Delta\hat{V}^A (\bm{k}',\bm{k})&=\frac{4\pi G}{\bm{q}^{2}}\Delta d_{1}^{A}(\bm{k}',\bm{k})+\frac{2\pi^{2}G^{2}}{|\bm{q}|}\Delta d_{2}^{A}(\bm{k}',\bm{k})+\mathcal{O}(G^{3}).
\end{align}
Tidal effects arise first at $\mathcal{O}(G^{2})$, which imposes
$\Delta d_{1}^{A}(\bm{k}^\prime,\bm{k})=0$.
Note also that the tidal effects allow for higher powers of $\bm{q}^2$ to contribute
classically to the potential at a given order in $G$.
This potential is computed from the EFT of ref.~\cite{Bern:2020buy}, whose action for spin-1/2 fermions is given by
\begin{align}\label{eq:EFT}
    S &= \int_{\bm{k}} \sum_{a=1,2} \psi^\dagger_a (-\bm{k}) \left( i\partial_t  - \sqrt{{\bm{k}}^2 + m_i^2}\right) \psi_a({\bm{k}}) \\
    &\qquad\qquad\qquad - \int_{\bm{k},\bm{k}'} \psi_1^\dagger(\bm{k}')\psi_2^{\dagger}(-\bm{k}')
    \left( \hat{V}(\bm{k}',\bm{k},\hat{\bm{S}}_a) + \Delta\hat{V}(\bm{k}',\bm{k},\hat{\bm{S}}_a) \right)
    \psi_1(\bm{k})\psi_2(-\bm{k}).
    \nonumber
\end{align}
The coefficients $\Delta d_{i}^{A}(\bm{k}',\bm{k})$ are then found by
matching the amplitudes from the above EFT with the amplitudes
derived from the tidal actions given in \cref{sec:opBasisQED,sec:opBasisGR}.

\begin{figure}
    \centering
    \includegraphics[scale=0.1]{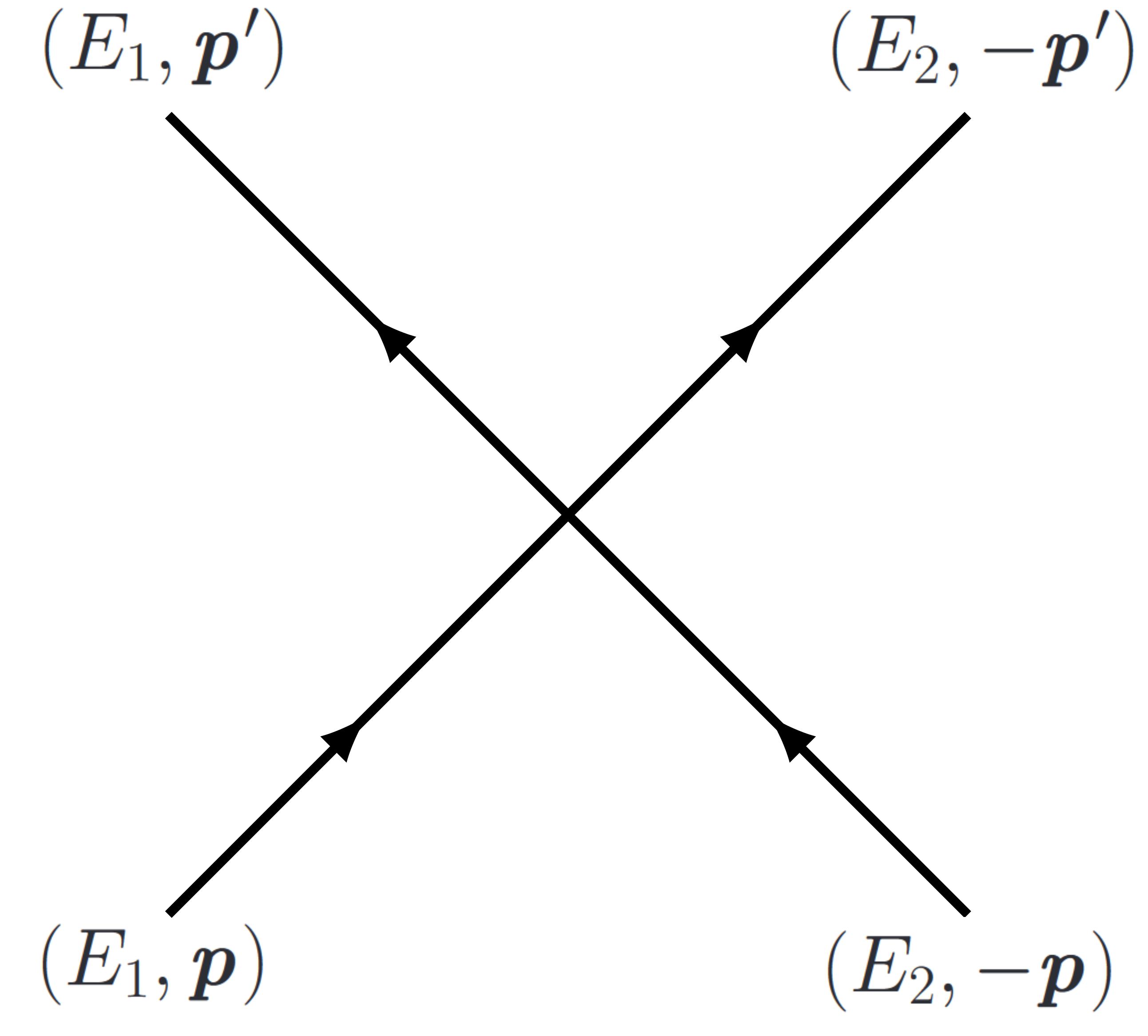}
    \caption{\label{fig:EFTDiag}The diagram encoding the leading tidal effect in the EFT in \cref{eq:EFT}.}
\end{figure}

We move now to the computation of the tidal amplitude from this EFT.
As noted above, $\Delta d_{1}^{A}(\bm{k}',\bm{k})=0$, which means the leading tidal potential is an
$\mathcal{O}(G^{2})$ quantity.
As we wish to match to the $\mathcal{O}(G^{2})$ amplitude from the full theory,
this means that we will only need to compute the tidal contribution
to the EFT amplitude from the tree-level diagram in \cref{fig:EFTDiag}.
This amplitude is simply
\begin{align}\label{eq:EFTAmp}
    \Delta\mathcal{M}^{\rm EFT}_{2}&=-\langle\bm{n}_{1}\bm{n}_{2}|\Delta \hat{V}(\bm{k}',\bm{k},\hat{\bm{S}}_{i})|\bm{n}_{1}\bm{n}_{2}\rangle\equiv-\Delta V(\bm{k}',\bm{k},\bm{S}_{i}).
\end{align}
The subscript $2$ on the left-hand side reminds us that this is an
amplitude at $\mathcal{O}(G^{2})$.
We can also expand the EFT amplitude in the basis of \cref{eq:SpinBasis}:
\begin{align}\label{eq:EFTAmpExpandedInSpin}
    \Delta\mathcal{M}^{\rm EFT}_{2}=\frac{2\pi^{2}G^{2}}{|\bm{q}|}&\left[\Delta a_{2}^{(0)}+\Delta a_{2}^{(1,1)}\bm{L}_{q}\cdot\bm{S}_{1}+a_{2}^{(1,2)}\bm{L}_{q}\cdot \bm{S}_{2}\right.\\
    &\quad\left.+\Delta a_{2}^{(2,1)}\bm{q}\cdot\bm{S}_{1}\bm{q}\cdot\bm{S}_{2}+\Delta a_{2}^{(2,2)}\bm{q}^{2}\bm{S}_{1}\cdot\bm{S}_{2}+\Delta a_{2}^{(2,3)}\bm{q}^{2}\bm{p}\cdot\bm{S}_{1}\bm{p}\cdot\bm{S}_{2}\right]. \nonumber
\end{align}
Combining this with \cref{eq:TidalSpinExp,eq:TidalPMExp,eq:EFTAmp} and
matching the coefficients on the different spin operators,
we arrive at the simple relation to leading-PM order in tidal effects:
\begin{align}\label{eq:EFTPotMatch}
    \Delta a_{2}^{A}&=-\Delta d_{2}^{A}.
\end{align}

Now, the EFT in \cref{eq:EFT} is the effective theory where the graviton
modes have been integrated out of the full theory.
As an effective version of the full theory, it must
produce the same low-energy amplitudes as the full theory:
\begin{align}\label{eq:AmpMatching}
    \Delta\mathcal{M}^{\rm EFT}_{2}&=\frac{\Delta\mathcal{M}^{s=1/2}_{2}}{4E_{1}E_{2}},
\end{align}
where we have non-relativistically normalized the full theory amplitude.
Note that since both of these amplitudes are the leading order where
tidal effects arise, there are no iteration terms to be matched.
Before we can match these two amplitudes, we must first express
the spinor products in the amplitude in \cref{eq:AmpCovCoeffsF}
in terms of the canonical spin vector.
The result of this is the mixing of the $\Delta a_{{\rm cov},2}^{A}$ coefficients, which has been worked out in ref.~\cite{Bern:2020buy}.\footnote{We thank Andr\'{e}s Luna for discussion on this point.}
The mixing derived there is slightly modified in our case, as our
spinors are dimensionful.
Accounting for this mixing, \cref{eq:AmpMatching} yields the matching conditions
\begin{align}
	\Delta a_{2}^{(0)}&= 4m_{1}m_{2}\Delta a_{{\rm cov},2}^{(0)}, \\
    \Delta a_{2}^{(1,1)}&= 2m_{2}\Delta a_{{\rm cov},2}^{(1,1)}-\frac{4m_{2}\Delta a_{{\rm cov},2}^{(0)}}{E(p_{1})+m_{1}},\\
	\Delta a_{2}^{(1,2)}&= 2m_{1}\Delta a_{{\rm cov},2}^{(1,2)}-\frac{4m_{1}\Delta a_{{\rm cov},2}^{(0)}}{E(p_{2})+m_{2}},\\
    \Delta a_{2}^{(2,1)} &= \Delta a_{{\rm cov},2}^{(2,1)}-\frac{2\bm{p}^{2}\Delta a_{{\rm cov},2}^{(1,1)}}{E(p_{2})+m_{2}}-\frac{2\bm{p}^{2}\Delta a_{{\rm cov},2}^{(1,2)}}{E(p_{1})+m_{1}}+\frac{4\bm{p}^{2}\Delta a_{{\rm cov},2}^{(0)}}{[E(p_{1})+m_{1}][E(p_{2})+m_{2}]},\\ 
    \Delta a_{2}^{(2,2)} &= \Delta a_{{\rm cov},2}^{(2,1)}+\frac{2\bm{p}^{2}\Delta a_{{\rm cov},2}^{(1,1)}}{E(p_{2})+m_{2}}+\frac{2\bm{p}^{2}\Delta a_{{\rm cov},2}^{(1,2)}}{E(p_{1})+m_{1}}-\frac{4\bm{p}^{2}\Delta a_{{\rm cov},2}^{(0)}}{[E(p_{1})+m_{1}][E(p_{2})+m_{2}]},\\ 
    \Delta a_{2}^{(2,3)} &= \Delta a_{{\rm cov},2}^{(2,1)}-\frac{2\Delta a_{{\rm cov},2}^{(1,1)}}{E(p_{2})+m_{2}}-\frac{2\Delta a_{{\rm cov},2}^{(1,2)}}{E(p_{1})+m_{1}}+\frac{4\Delta a_{{\rm cov},2}^{(0)}}{[E(p_{1})+m_{1}][E(p_{2})+m_{2}]}.
\end{align}
Combining this with \cref{eq:EFTAmp,eq:EFTAmpExpandedInSpin,eq:EFTPotMatch}, the momentum-space
potential at $\mathcal{O}(G^{2})$ can be written as
\begin{align}
    \Delta V(\bm{k}',\bm{k},\bm{S}_{a})&=-\frac{2\pi^{2}m_{2}^{2}G^{2}}{|\bm{q}|}\sum_{j=0}^{\infty}\left(\frac{\bm{q}^{2}}{2}\right)^{j+2}\sum_{A}\Delta a_{2,j}^{A}(\omega)\mathbb{O}^{A},
\end{align}
where the $\Delta a_{2,j}^{A}(\omega)$ are defined through
\begin{align}
    \Delta a_{2}^{A}(\omega)\equiv m_{2}^{2}\sum_{j=0}^{\infty}\left(\frac{\bm{q}^{2}}{2}\right)^{j+2}\Delta a_{2,j}^{A}(\omega).
\end{align}
The potential can be expressed in position space by simply
Fourier transforming, giving (see \Cref{app:fourierInt} for the relevant integrals)
\begin{align}
    \Delta V(\bm{r},\bm{p},\bm{S}_a)             &=-\frac{\sqrt{\pi}m_{2}^{2}G^{2}}{2}\sum_{j=0}^{\infty}
    \left(\frac{2}{\bm{r}^{2}}\right)^{j+3}\Gamma(j+3)\left[\frac{1}{\Gamma(-j-3/2)}\Delta a_{2,j}^{(0)}(\omega)\, \mathbb{I} \right.\notag \\
	& - \left(\frac{2}{\bm{r}^{2}}\right)\frac{j+3}{\Gamma(-j-3/2)}\left(\Delta a_{2,j}^{(1,1)}(\omega)\bm{L}\cdot \bm{S}_1+\Delta a_{2,j}^{(1,2)}(\omega)\bm{L}\cdot \bm{S}_2\right)\notag \\ 
	&-2(j+4)\left(\frac{2}{\bm{r}^{2}}\right)\frac{1}{\bm{r}^{2}}\frac{j+3}{\Gamma(-j-3/2)}\Delta a_{2,j}^{(2,1)}(\omega)\bm{r}\cdot\bm{S}_1 \bm{r}\cdot\bm{S}_2 \notag \\
	&+\left(\frac{2}{\bm{r}^{2}}\right)(j+3)\left(\frac{1}{\Gamma(-j-3/2)}\Delta a_{2,j}^{(2,1)}(\omega) + \frac{2}{\Gamma(-j-5/2)}\Delta a_{2,j}^{(2,2)}(\omega) \right) \bm{S}_1\cdot \bm{S}_2 \notag \\
	&\left. + 2\left(\frac{2}{\bm{r}^{2}}\right)\frac{j+3}{\Gamma(-j-5/2)}\Delta a_{2,j}^{(2,3)}(\omega) \bm{p}\cdot \bm{S}_1 \bm{p}\cdot \bm{S}_2\right],
\end{align}
where $\bm{L}\equiv\bm{r}\times\bm{p}$ is the angular momentum.
The spin-monopole portion of this potential agrees with that found in ref.~\cite{Haddad:2020que}
after performing the matching in \cref{eq:MonopoleAmpMatch1,eq:MonopoleAmpMatch2}

\section{Classical observables}
\label{sec:impulse}

The scattering amplitude can be related to various classical observables.
For example, refs.~\cite{Kosower:2018adc,Maybee:2019jus} derived direct relations between the
scattering amplitude and the linear and angular impulses. 
In this section, we will use the tidally deformed scattering amplitudes
for spin-1/2 particles, \cref{eq:EMampResult,eq:GRampResult}, to find the tidal contributions to these quantities.
We remark that the $\Delta$ prefixes in the remainder of this section denote
changes in the linear and angular momenta.

\subsection{Linear impulse}

Following ref.~\cite{Kosower:2018adc}, the leading term for the classical linear impulse is\footnote{We leave the factors of $\hbar$ implicit.}
\begin{align}\label{eq:KMOCLinImp}
	\Delta p_1^{\mu} =  \llangle[\Bigg]  i  \int \hat d^4  q \hat \delta(2p_1\cdot  q) \hat \delta(2p_2\cdot  q) e^{-ib\cdot  q}
	q^\mu \mathcal{A}(p_1,p_2\rightarrow p_1+ q,p_2- q)\rrangle[\Bigg] ,
\end{align}
where $b^{\mu}$ is the impact parameter.
Since we are dealing with the leading-PM tidal amplitudes, we do not need to include the second contribution to the linear impulse, i.e. eq.~(3.30) in ref.~\cite{Kosower:2018adc}.
The effect of the double angle brackets is to take the expectation values for the spins, as well as to impose the replacement
$p_{i}\rightarrow m_{i}v_{i}$ on the initial momenta.
As we have already written the amplitudes by expressing the momenta in this way,
the task is reduced to the computation of integrals of the form in \cref{eq:ClassInt}.
We present all computational details in \cref{app:fourierInt} and write here just the results.

Before computing the impulse, we highlight that we will work with the covariant
spin vector in this section.
This changes the way the spinor products $\mathcal{U}_{i}$ are treated compared
to \Cref{sec:Hamiltonian}.
First, note that the spinors
in the amplitudes of \cref{eq:EMampResult,eq:GRampResult}
are normalized such that $\bar{u}(p)u(p)=2m$.
Moreover, as the momenta of the final states differ from those
of the initial states by momenta of order $\hbar$,
we can expand the final state spinors in powers of $\hbar$:
\begin{align}\label{eq:SpinorExp}
    \bar{u}(p\pm \hbar\bar{q})&=\bar{u}(p)+\mathcal{O}(q).
\end{align}
In fact, when working with the covariant spin vector, 
this $\mathcal{O}(q)$ correction is an infinitesimal Lorentz boost of the spinor \cite{Maybee:2019jus}.
This applies just as well to the heavy spinors, so,
to leading order in $q$, the spinor products are $\mathcal{U}_{i}=2m_{i}+\mathcal{O}(q^{2})$.

We begin with the linear impulse in the gravitational case.
Plugging the tidal contribution to the amplitude, \cref{eq:GRampResult}, into the impulse formula we have that\footnote{Note that \cref{eq:KMOCLinImp} gives the impulse for the particle that absorbs the transfer momentum.
In our case, particle 1 is the emitting particle, so we must evaluate the integrals at $-b^{\mu}$.}
\begin{align}
	\Delta &p^{\mu}_{1,\rm GR} 
         = i G^2 m_2^2  \sum_{j=0}^{\infty}
        \left[ 
		4m_1 m_2 G^{(0)}_j I_{j+2}^{\mu}(-b) \right.\nonumber \\
	  & -2i\omega m_{1}m_{2}\epsilon_{\rho\nu\alpha\beta}v_{1}^{\nu} v_{2}^{\alpha}\left( m_2  \langle S_{1}^{\beta}\rangle  G_{j}^{(1,1)} - m_1 \langle S_{2}^{\beta} \rangle  G_{j}^{(1,2)}\right) I_{j+2}^{\mu\rho}(-b) \nonumber \\ &
	  \left.+ \left( \langle S_{1\nu}\rangle \langle S_{2\rho} \rangle G_j^{(2,1)}- \eta_{\nu\rho} \langle S_1\rangle \cdot \langle S_2\rangle G_j^{(2,2)}
		  + \omega\eta_{\nu\rho} \left(v_2\cdot \langle S_1\rangle\right)\left(  v_1\cdot \langle S_2\rangle\right)G_j^{(2,3)}\right) I_{j+2}^{\mu\nu\rho}(-b)
        \right] .
\end{align}
By using the results for the integrals in \cref{eq:Irank1result,eq:Irank2result,eq:Irank3result},
we end up with
\begin{align}
	\label{eq:impulseGRresult}
	\Delta p^{\mu}_{1,\rm GR} 
	=&\frac{-\pi   G^2 m_2}{8 m_1  \sqrt{\omega^2-1} }
	 \sum_{j=0}^{\infty} \left(-\frac{2}{b^2}\right)^{j+3}
	 \frac{\Gamma[7/2+j]}{\Gamma[-3/2-j]} 
        \left[ 
		-4m_1 m_2 G^{(0)}_j\frac{b^\mu}{|\bm{b}|} \right. \nonumber\\
	 &\left.+ 2\omega m_1 m_2^2 \epsilon_{\rho\nu\alpha\beta}v_{2}^{\nu} v_{1}^{\alpha} \langle S_{1}^{\beta}\rangle  G_{j}^{(1,1)}
	\frac{1}{|\bm{b}|} 
	\left(
		(7+2j)\frac{b^\mu b^\rho }{b^2}
	- \Pi^{\mu\rho}
	\right)
	\right.\nonumber \\
	 &+ 2\omega m_1^2  m_2 \epsilon_{\rho\nu\alpha\beta} v_{1}^{\nu} v_{2}^{\alpha} \langle S_{2}^{\beta} \rangle  G_{j}^{(1,2)} 
	\frac{1}{|\bm{b}|} 
	\left(
		(7+2j)\frac{b^\mu b^\rho }{b^2}
	- \Pi^{\mu\rho}
	\right)
	\nonumber \\ &
	  - \left[ \langle S_{1\nu}\rangle \langle S_{2\rho} \rangle G_j^{(2,1)}	
 - \eta_{\nu\rho}\left(   \langle S_1\rangle \cdot \langle S_2\rangle \right)G_j^{(2,2)}	
	  + \omega \eta_{\nu\rho}\left(v_2\cdot \langle S_1\rangle\right)\left(  v_1\cdot \langle S_2\rangle\right)G_j^{(2,3)}  \right]
        \nonumber \\
	&\times \left.
	\left(
	(9/2+j)\frac{b^\mu b^\nu b^\rho}{b^2}
	-
\frac{3}{2}b^{(\mu}\Pi^{\nu\rho)}\right)
	\frac{2}{|\bm{b}|}
\left(-\frac{2}{b^2}\right)  (7/2+j) \right] ,
\end{align}
where $\Pi^{\mu\nu}$ is defined in \cref{eq:PiProjector}.
In the electromagnetic case, the impulse is
\begin{align}
	\Delta p^{\mu}_{1,\rm EM} 
	=&\frac{-  e^2}{8 \pi m_1 m_2  \sqrt{\omega^2-1} }
	 \sum_{j=0}^{\infty} \left(-\frac{2}{b^2}\right)^{j+2}
	 \frac{\Gamma[5/2+j]}{\Gamma[-1/2-j]} 
        \left[ 
		-4m_1 m_2 F^{(0)}_j\frac{b^\mu}{|\bm{b}|} \right. \nonumber\\
	 &\left.+ 2\omega m_1 m_2^2 \epsilon_{\rho\nu\alpha\beta}v_{2}^{\nu} v_{1}^{\alpha} \langle S_{1}^{\beta}\rangle  F_{j}^{(1,1)} 
	\frac{1}{|\bm{b}|} 
	\left(
		(5+2j)\frac{b^\mu b^\rho }{b^2}
	- \Pi^{\mu\rho}
	\right)
	\right.\nonumber \\
	 &+ 2\omega m_1^2  m_2 \epsilon_{\rho\nu\alpha\beta} v_{1}^{\nu} v_{2}^{\alpha} \langle S_{2}^{\beta} \rangle  F_{j}^{(1,2)} 
	\frac{1}{|\bm{b}|} 
	\left(
		(5+2j)\frac{b^\mu b^\rho }{b^2}
	- \Pi^{\mu\rho}
	\right)
	\nonumber \\ &
	  - \left[\langle S_{1\nu}\rangle \langle S_{2\rho} \rangle F_j^{(2,1)}	
 - \eta_{\nu\rho}\left( \langle S_1\rangle \cdot \langle S_2\rangle \right)F_j^{(2,2)}	
	  + \omega \eta_{\nu\rho}\left(v_2\cdot \langle S_1\rangle\right)\left(  v_1\cdot \langle S_2\rangle\right)F_j^{(2,3)}  \right]
        \nonumber \\
	&\times \left.
	\left(
	(7/2+j)\frac{b^\mu b^\nu b^\rho}{b^2}
	-
\frac{3}{2}b^{(\mu}\Pi^{\nu\rho)}\right)
	\frac{2}{|\bm{b}|}
\left(-\frac{2}{b^2}\right)  (5/2+j) \right] .
\end{align}
Note that the electromagnetic result can be obtained from the gravitational
result using the following replacements:
\begin{align}\label{eq:GRtoEMObsMap}
	G^2 m_2^2 &\rightarrow e^2/\pi^2 , \\
	G_{j}^{A} &\rightarrow F_j^{A} , \\
	j &\rightarrow j-1 .
\end{align}
The last of these replacements is not applied to the indices of the form factors.

\subsection{Angular impulse}

We turn now to the determination of the angular impulse.
The angular impulse for the absorbing particle
is related to the amplitude through \cite{Maybee:2019jus}\footnote{Again, we leave the factors of $\hbar$ implicit.}
\begin{align}
	\Delta S_1^{\mu} =
	\llangle[\Bigg]
	i \int \hat d^4  q \hat \delta(2p_1\cdot  q)\hat \delta(2p_2\cdot  q)
	e^{-ib\cdot  q}
	\left(
		-  \frac{p_1^\mu}{m_1^2}  q\cdot S_1(p_1)\mathcal{A}(q)
		+ \left[
			S_1^\mu(p_1),\mathcal{A}(q)
		\right]
	\right)
	\rrangle[\Bigg],
\end{align}
where $p_{1}^{\mu}$ is the initial momentum of the absorbing particle.
We don't need the second contribution to the impulse, eq.~(3.22) in ref.~\cite{Maybee:2019jus},
because we are calculating the leading-PM tidal contribution.
For the spin-1/2 amplitudes we are considering, this formula
will produce terms of $\mathcal{O}(S_{i}^{2})$.
We ignore such contributions
since one must consider spin-1 scattering to obtain all
information at this spin order.

We compute each term individually for particle 1, which is the emitting particle in our setup.
The first term is
\begin{align}
	\label{eq:spinkick1}
    \llangle[\Bigg]
	i \int \hat d^4  q \hat \delta(2p_1\cdot  q)\hat \delta(2p_2\cdot  q)
	e^{ib\cdot  q}
		  \frac{p_1^\mu}{m_1^2}  q\cdot S_1(p_1)\mathcal{A}_{2}(q)
	\rrangle[\Bigg]&=-\frac{v_{1}^{\mu}}{m_{1}}\Delta p_1\cdot \langle S_{1}\rangle,
\end{align}
which follows since, to the order we are working, there is only
one contribution to the linear impulse.
This is true for both the electromagnetic and gravitational cases.
Writing this explicitly for the case of gravity,
\begin{align}
    - \frac{v_1^\mu}{m_1}&\Delta p_{1}\cdot \langle S_{1}\rangle= \frac{v_{1}^{\mu}\pi G^2 m_2^2}{4m_1 |\mathbf{b}|\sqrt{\omega^2-1} }
	 \sum_{j=0}^{\infty} \left(-\frac{2}{b^2}\right)^{j+3}
	 \frac{\Gamma[7/2+j]}{\Gamma[-3/2-j]} \\
        &\times\left[- 
		2G^{(0)}_j b^\lambda+ \omega m_1 \epsilon_{\rho\nu\alpha\beta} v_{1}^{\nu} v_{2}^{\alpha} \langle S_{2}^{\beta} \rangle  G_{j}^{(1,2)}\left((7+2j)\frac{b^\lambda b^\rho }{b^2}-\Pi^{\lambda\rho}\right)\right]\langle S_{1\lambda}\rangle+\mathcal{O}(\langle S_{1}\rangle^{2}).\notag
\end{align}
To compute the commutator term, we need the following commutator \cite{Maybee:2019jus}:
\begin{align}
    [S_{i}^{\mu},S_{j}^{\nu}]&=-\delta_{ij}\frac{i}{m_{i}}\epsilon^{\mu\nu\rho\sigma}S_{i\rho}p_{i\sigma}.
\end{align}
With this in hand, the commutator term for gravity is
\begin{align}
	\label{eq:spinkickGRresult}
    &\llangle[\Bigg] 
	i \int \hat d^4 q \hat \delta(2p_1\cdot q)\hat \delta(2p_2\cdot q)
	e^{ib\cdot q}
	 \left[
			S_1^\mu(p_1),\mathcal{M}_{2}(q)
		\right]
	\rrangle[\Bigg]\notag \\
    &\qquad= i G^2 m_2^2 \sum_{j=0}^{\infty}
        \left[ 
		 -2\omega  m_1 m_2^2 G_{j}^{(1,1)}
		 \left[(v_2^\mu- v_1^\mu \omega) I_{j+2}^{\alpha}  - I_{j+2}^{\mu}   v_{2}^{\alpha}\right]   \langle S_{1\alpha}\rangle
	\right.\nonumber \\& \qquad
	\left.-i {\epsilon^{\mu}}_{\nu\lambda\tau} v_1^{\tau} \left[\delta^{\nu}_{\alpha}\delta^{\eta}_{\beta} G_j^{(2,1)}
 - \eta_{\alpha\beta} \eta^{\nu\eta} G_j^{(2,2)}
	+\eta_{\alpha\beta} \omega  v_{2}^{\nu}  v_1^\eta G_j^{(2,3)} \right]I_{j+2}^{\alpha\beta} \langle S_{1}^{\lambda}\rangle \langle S_{2\eta} \rangle 
	\right]+\mathcal{O}(\langle S_{1}\rangle^{2})  .
\end{align}
We simply add \cref{eq:spinkick1} and 
\cref{eq:spinkickGRresult} to get the full angular impulse of particle 1
in the gravitational case.
Again, the map in \cref{eq:GRtoEMObsMap} can be applied to the gravitational impulse to obtain the electromagnetic result.
The same calculation can be performed for particle 2, but we don't give the
result here as it is almost identical to the calculation for particle 1.

\section{Eikonal phase}\label{sec:eikonal}

The eikonal phase provides an alternative means for extracting physical observables from the classical portion of scattering amplitudes,
and has been successfully applied to systems involving low spins up to $\mathcal{O}(G^{2})$ \cite{Guevara:2018wpp,Bern:2020buy}.
Relations between classical observables and the eikonal phase were proposed to all perturbative orders
in ref.~\cite{Bern:2020buy};
however, as we are working with tidal effects at the leading-PM order, we only need these relations
to leading order.
For the linear impulse and the spin kick, they are
\begin{align}\label{eq:EikonalObs}
	\Delta \bm{p}_\perp = \nabla_{\bm{b}} \chi 
	\quad \text{and} \quad
	\Delta \bm{S}_a^i = - \epsilon^{ijk} \frac{\partial \chi}{\partial \bm{S}_a^j} \bm{S}_a^k ,
\end{align}
where $\chi$ is the eikonal phase, $\Delta$ indicates changes in momentum or spin, and $a=1,2$ labels the particles and is not summed over.
We use Latin letters from the middle of the alphabet to denote spatial indices,
which are raised and lowered with the Euclidean metric.
The former relation yields the linear impulse in the plane perpendicular to the momentum at negative infinity.
Choosing kinematics such that this momentum is oriented along the $z$-axis,
the impulse in the parallel direction is obtained through energy conservation: $\Delta p_z = - (\Delta \bm{p})^2/2|\bm{p}|$.
Further to the results in \Cref{sec:impulse}, the eikonal phase will allow us to compute the spin kick and the scattering angle for aligned spins.
First, we must compute the eikonal phase.

The eikonal phase $\chi = \chi_1 + \chi_2 + \dots$ is defined as the Fourier transform of the (relativistically normalized) amplitude in the perpendicular plane described above,
with the subscripts denoting contributions from the corresponding order in the coupling constant:
\begin{align}
	\chi_i =& \frac{1}{4m_1 m_2 \sqrt{\omega^2-1}}\int \frac{d^{2-2\epsilon}\bm{q}}{(2\pi)^{2-2\epsilon}}
		e^{-i \bm{q}\cdot\bm{b}} \mathcal{M}^\prime_{i}(\bm{q}) ,
\end{align}
where the prime on the amplitude indicates that we ignore iteration pieces.
We will use $\Delta\chi$ to denote tidal contributions to the eikonal phase.
The leading-PM contributions from the tidal deformations originate from a one-loop amplitude, so $\Delta\chi_{1}=0$ and
we will calculate the $\Delta \chi_2$ part of the eikonal phase.
This involves integrals of the type
\begin{align}
\int \frac{d^2\bm{q}}{(2\pi)^2} e^{-i\bm{q}\cdot \bm{b}}
\frac{\pi^{2}}{|\bm{q}|} \left(\frac{\bm{q}^2}{2}\right)^{j+2}
\hat{\mathbb{O}}^A(\bm{p},\bm{q},\bm{S}_a)  
=\hat{\mathcal{O}}^A(\bm{p},\nabla_{\bm{b}},\bm{S}_a)   I_{j+2} (\bm{b})\,\frac{\mathcal{N}}{4},
\end{align}
where we have shifted $\bm{q}\rightarrow-\bm{q}$ on the right-hand side to eliminate relative signs with the integrals in \cref{app:fourierInt};
$\mathcal{N}$ and $I_{j+2}$ are given there.
The classical spin structures $\hat{\mathbb{O}}^{A}$ are listed in \cref{eq:SpinBasis}. They have been removed
from the integral by re-expressing them as differential operators in impact-parameter space acting on the integral, yielding
\begin{align}\label{eq:SpinBasisImpactPar}
\hat{\mathcal{O}}^{(0)} = \mathbb{I}, \quad
\hat{\mathcal{O}}^{(1,1)} &= -( \bm{S}_1 \times \bm{p})\cdot \nabla_{\bm{b}}, \quad \hat{\mathcal{O}}^{(1,2)} =  -(\bm{S}_2 \times \bm{p})\cdot \nabla_{\bm{b}},\notag \\
\hat{\mathcal{O}}^{(2,1)} = -(\bm{S}_1 \cdot \nabla_{\bm{b}}) (\bm{S}_2 \cdot \nabla_{\bm{b}}), \quad
\hat{\mathcal{O}}^{(2,2)} &= -(\bm{S}_1 \cdot \bm{S}_2)\, \nabla_{\bm{b}}^2, \quad
\hat{\mathcal{O}}^{(2,3)} = -(\bm{p} \cdot \bm{S}_1)(\bm{p} \cdot \bm{S}_2) \nabla_{\bm{b}}^2.
\end{align}
This allows us to write the eikonal phase $\Delta \chi_2$ compactly as,
\begin{align}
\label{eq:EikonalCompact}
\Delta\chi_2 &= 2 G^2 m_2^2 \sum_{j=0}^{\infty}
\hat{\mathcal{K}}_j(\omega,\bm{p},\nabla_{\bm{b}},\bm{S}_a)\,  I_{j+2}(\bm{b}),\quad \hat{\mathcal{K}}_j(\omega,\bm{p},\nabla_{\bm{b}},\bm{S}_a)\equiv 4E_{1}E_{2}\sum_A \Delta b^A_{2,j}(\omega)\hat{\mathcal{O}}^A,
\end{align}
where the dependence on each quantity has been written explicitly. 
The function $\Delta b_{2,j}^{A}(\omega)$ is either equal to $m^{A}\Delta a_{{\rm cov},2,j}^{A}(\omega)$ (no sum over $A$) or
$\Delta a_{2,j}^{A}(\omega)$, depending on whether we work with the covariant or the canonical spin.
The constant $m^{A}$ accounts for the normalization of the Dirac spinors, and is given by
\begin{align}
    m^{A}&=\begin{cases}
    4m_{1}m_{2}, & A=(0), \\
    2m_{2}, & A=(1,1), \\
    2m_{1}, & A=(1,2), \\
    1, & {\rm otherwise.}
    \end{cases}
\end{align}
We refer to the operator $\hat{\mathcal{K}}_{j}$ as the eikonal operator.

We can now substitute this form of the eikonal phase into \cref{eq:EikonalObs} to obtain the impulse and
spin kick.
For the impulse, the derivative with respect to the impact parameter commutes with the eikonal operator
and only acts on the integral.
By contrast, for the spin kick, the derivative with respect to the spin vector only acts on the eikonal operator:\footnote{The sign for the linear impulse here is because we compute the impulse for the emitting particle.}
\begin{align}
\Delta \bm{p}_\perp &=-2 G^2 m_2^2 \sum_{j=0}^{\infty}\hat{\mathcal{K}}_j(\omega,\bm{p},\nabla_{\bm{b}},\bm{S}_a) \left(\nabla_{\bm{b}} I_{j+2}(\bm{b})\right) ,
\\
\Delta \bm{S}_a &= -2 G^2 m_2^2 \sum_{j=0}^{\infty}\left(\frac{\partial \hat{\mathcal{K}}_j(\omega,\bm{p},\nabla_{\bm{b}},\bm{S}_a)}{\partial\bm{S}_a} \times \bm{S}_a\right) I_{j+2}(\bm{b}).
\end{align}
Setting $\Delta b_{2,j}^{A}(\omega)=m^{A}\Delta a_{{\rm cov},2,j}^{A}(\omega)$, the linear impulse calculated from the eikonal phase agrees with \cref{eq:impulseGRresult}.
We find the spin kick for particle 1 to be
\begin{align}
    \Delta &\bm{S}_{1}^k=- 8E_{1}E_{2}G^2m_2^2 \sum_{j=0}^{\infty} \left\{\,-i\Delta b_{2,j}^{(1,1)}[-\bm{p}^k \bm{S}_{1}^{m} +(\bm{p}\cdot \bm{S}_1)\,\delta^{km}]I_{j+2}^m\right. \nonumber\\
    &\left.+\left[-\Delta b_{2,j}^{(2,1)}(\varepsilon^{k\ell m} \bm{S}_{1}^{\ell}) \bm{S}_{2}^{n} - \Delta b_{2,j}^{(2,2)} (\bm{S}_1 \times \bm{S}_2)^k \delta^{mn}+\Delta b_{2,j}^{(2,3)}(\bm{p} \times \bm{S}_1)^k(\bm{p} \cdot \bm{S}_2)\delta^{mn} \right]I^{mn}_{j+2}\right\}.
\end{align}
This expression for the spin kick is valid in the center-of-mass frame.
Computing the same quantity in the rest frame of particle 1,
we find agreement with the commutator portion of the angular impulse, \cref{eq:spinkickGRresult}.

From the eikonal phase it is also possible to obtain the scattering angle. 
In the case of two non-rotating bodies, the dynamics are constrained to a plane, so a unique scattering angle can be defined.
Spinning particles, however, introduce precession effects, which are described by an additional angle. 
This additional complication can be ignored in the special case of aligned spins,
which again restricts motion to a plane.
We will calculate the unique scattering angle in this special case, where the spin structures are
\begin{align}
    \bm{S}_{1}\cdot \bm{S}_{2}=|\bm{S}_{1}||\bm{S}_{2}|,\quad \bm{b}\cdot \bm{S}_{i}=0,\quad \bm{p}\cdot \bm{S}_{i}=0.
\end{align}
This scattering angle can be related to the eikonal phase using the stationary phase approximation~\cite{Amati:1987uf}:
\begin{align}
2\sin \frac{\theta}{2}\approx\theta = -\frac{1}{|\bm{p}|}\frac{\partial}{\partial|\bm{b}|}\chi(\omega,\bm{b}).
\end{align}
For the calculation of the scattering angle,
we will use the eikonal phase in terms of the canonical spin,
as it is to be compared to the scattering angle derived from canonical equations of motion. 
Thus, $\Delta b_{2,j}^{A}(\omega)=\Delta a_{2,j}^{A}(\omega)$.

As in the case of the linear impulse, the derivative with respect to $|\bm{b}|$ acts only on the integral.
Using the relation
\begin{align}
\frac{\partial}{\partial |\bm{b}|} I_{j+2}(\bm{b}) = \frac{2|\bm{b}|}{(5+2j)}I_{j+3}(\bm{b}),
\end{align}
we can write the scattering angle generally as
\begin{align}
\Delta \theta = -\frac{2G^2 m_2^2}{|\bm{p}|} \sum_{j=0}^{\infty}
\frac{2|\bm{b}|}{(5+2j)} \hat{\mathcal{K}}_j(\omega,\bm{p},\nabla_{\bm{b}},\bm{S}_a)\,  I_{j+3}(\bm{b}).
\end{align}
Using the action of each differential operator on the integral (\cref{eq:EikonalOpTerms}), the tidal corrections to the 2PM aligned-spin scattering angle are
\begin{align}
\Delta \theta_2 
&= \frac{\pi G^2 m_2 E_{1}E_{2}}{|\bm{p}|m_1 \sqrt{\omega^2-1}} \sum_{j=0}^{\infty}
\frac{\Gamma[9/2+j]}{\Gamma[-3/2-j]}
\left(\frac{2}{\bm{b}^2}\right)^{j+4} \nonumber \\
&\times\left[\Delta a_{2,j}^{(0)} \frac{\bm{b}^2}{(7+2j)}+
\Delta a_{2,j}^{(1,1)} (\bm{S}_1 \times \bm{p})\cdot \bm{b}+ 
  \Delta a_{2,j}^{(1,2)} (\bm{S}_2 \times \bm{p})\cdot\bm{b}\right.\notag\\
& \qquad\left. + \left(\Delta a_{2,j}^{(2,1)}
-\left(5+2j\right)\Delta a_{2,j}^{(2,2)}\right)|\bm{S}_{1}||\bm{S}_{2}| 
\right].
\end{align}
The spin-monopole portion of this is in agreement with that in ref.~\cite{Haddad:2020que},
upon applying the matching conditions in \cref{eq:MonopoleAmpMatch1,eq:MonopoleAmpMatch2}.

\section{Conclusion}\label{sec:conclusion}

As the recent burst in activity suggests, quantum-field-theoretic techniques are well suited for studying tidal deformations, where
the tidal effects are characterized by higher-dimensional operators.
A full classification of tidal operators relevant for tidally-deformed spinless objects at the one-loop level
was presented in ref.~\cite{Haddad:2020que},
and in this paper we have extended this analysis to include effects at linear order in the spin of the deformed object.
As in ref.~\cite{Haddad:2020que}, the starting point was the Hilbert series, which counts the number of independent operators
-- equivalently, the number of independent amplitudes --
for a given field content and number of covariant derivatives.
Using this as a guide, we wrote down both the amplitude basis and the operator basis for a spin-1/2 particle
coupled to photons or gravitons through at most two photon field strengths/Weyl tensors.
These operator bases represent the full set of operators coupling two spinors
to two photon field strengths or Weyl tensors,
describing the complete set of finite-size contributions at one loop.

Employing traditional Feynman diagrammatic methods, we used these actions to
calculate the one-loop amplitudes -- corresponding to the leading-PM order in the case of gravity --
for these finite-size effects.
We find that the spin-multipole universality for long-range classical effects observed
in refs.~\cite{Holstein:2008sw,Holstein:2008sx} extends to tidally deformed systems;
the spin-monopole portions of the amplitudes calculated here are in agreement with those found
in ref.~\cite{Haddad:2020que}.
For general Wilson coefficients, the finite-size contributions 
to the amplitudes break the observed correspondence between the $q\cdot S_{1}\,q\cdot S_{2}$ and $q^{2}S_{1}\cdot S_{2}$ terms
in the point-particle case.

We then extracted various classical quantities from these amplitudes.
First, we extended the EFT matching formalism of ref.~\cite{Bern:2020buy} to include 
tidal effects, and subsequently used this formalism to derive the tidal
corrections to the conservative gravitational Hamiltonian at leading-PM order.
We then derived the finite-size contributions to the electromagnetic and gravitational
linear and angular impulse.
The linear impulse was computed in two ways for the gravitational case, producing the same result:
it was first calculated using the formalism of ref.~\cite{Kosower:2018adc},
then through application of the eikonal phase \cite{Bern:2020buy}.
The angular impulse was computed using the method of ref.~\cite{Maybee:2019jus}.
A portion of this result was corroborated by the extraction of the spin kick from the
eikonal phase; when computing this quantity in the rest frame of particle 1,
it agrees with the commutator contribution to the angular impulse.
Finally, the eikonal phase allowed us to derive the scattering angle in the case of aligned spins.

In the interest of describing real macroscopic systems, one must account for
finite-size effects at arbitrary orders in the spin vector.
We have demonstrated that EFT techniques such as the
Hilbert series and the construction of on-shell helicity amplitudes
are suitable for combining spin and finite-size effects.
These techniques can yet be extended to higher spins, but this is outside the scope of this paper.

The description of finte-size effects for spinning particles also opens the door for the study of
the entanglement entropy generated in the scattering of tidally deformed objects.
A similar analysis to that performed in ref.~\cite{Aoude:2020mlg} can be applied to 
the systems described here, potentially shedding some 
light on the values of tidal Love numbers for
Kerr black holes in a general (but weak) gravitational environment.

Finally, it would be helpful to understand links between the
tidal action here and worldline actions
describing tidal effects at linear order in the angular momentum, 
perhaps by taking the heavy or non-relativistic limits
of the quantum action in \cref{eq:GRAction}.
The action in \cref{eq:GRAction} is the most general, non-redundant action
describing parity-even four-point contact terms, but not all operators necessarily need to
contribute classical effects;
for example, we found that certain operators enter only at subleading orders
at the one-loop level.
A matching to classical quantities is then essential for 
determining the subset of operators that do indeed contribute classically.

\acknowledgments

We thank Clifford Cheung, Andr\'{e}s Luna, Ben Maybee, and Julio Parra-Martinez for related discussions.
This project has received funding from the European Union's Horizon 2020
research and innovation programme under the Marie Sk\l{}odowska-Curie grant
agreement No. 764850 "SAGEX" and by the F.R.S.-FNRS with the EOS - be.h project n. 30820817.
A.H. is supported by the DOE under grant no. DE-SC0011632 and by the Walter Burke Institute for Theoretical Physics.

\appendix

\section{Loop integrals}\label{sec:integral}

For the calculation of tidal effects at one loop, we only need to evaluate the triangle integral,
but we need to do so for arbitrary even rank,
\begin{align}
	\mathcal{I}_\triangleleft^{\mu_1\dots\mu_{2k}} \equiv \int \frac{d^4l}{(2\pi)^4} \frac{l^{\mu_1\dots\mu_{2k}}}{l^2 (l+q)^2 [-v_2\cdot l + i\epsilon]} .
\end{align}
The Passarino-Veltman reduction \cite{PASSARINO1979151} allows one to solve this
for any rank in terms of the scalar triangle integral\footnote{More precisely, the Passarino-Veltman reduction expresses the rank-$2k$ integral in terms of scalar triangle, bubble, and sunset integrals, but neither of the latter two contributes classical information.}
\begin{align}
	\mathcal{I}_\triangleleft \equiv \int \frac{d^4l}{(2\pi)^4} \frac{1}{l^2 (l+q)^2 [-v_2\cdot l + i\epsilon]} =
	-\frac{i S}{16\pi^2},
\end{align}
where $S\equiv \pi^2/\sqrt{-q^2}$, but this becomes cumbersome for high ranks.

Fortunately, the portions of the integrals needed for our purposes are only the leading-in-$\hbar$ pieces.
When the tensor integral is dotted into $v_{1\mu_1\dots\mu_{2k}}$, ref.~\cite{Haddad:2020que} found by explicit calculation up to rank $2k=10$ that
this leading term is
\begin{align}\label{eq:SpinlessTriIntGenK}
	v_{1\mu_1\dots\mu_{2k}}\mathcal{I}_\triangleleft^{\mu_1\dots\mu_{2k}} \equiv \int \frac{d^4l}{(2\pi)^4} \frac{(v_1\cdot l)^{2k}}{l^2 (l+q)^2 [-v_2\cdot l + i\epsilon]}
	= \frac{\left(\frac{1}{2}\right)_{k}}{4^{k}(1)_k}(\omega^2-1)^{k}q^{2k} \mathcal{I}_\triangleleft + \mathcal{O}(q^{2k}).
\end{align}
This formula was proven for general $k$ in ref.~\cite{Bern:2020uwk} by finding the residue of the matter pole and then calculating the remaining three-dimensional integral using a known expression \cite{Smirnov:2012gma}.

When calculating the leading-PM amplitudes for spinning tidal effects, we also need five additional
integrals:
\begin{align}
	(\bar u_2 q^\alpha \sigma_{\alpha\mu_1}u_1) v_{1\mu_2\dots\mu_{2k}}\mathcal{I}_\triangleleft^{\mu_1\dots\mu_{2k}},\label{eq:Spin1Int} &\\
	(\bar u_4 q^\alpha \sigma_{\alpha\mu_1}u_3) v_{1\mu_2\dots\mu_{2k}}\mathcal{I}_\triangleleft^{\mu_1\dots\mu_{2k}},\label{eq:Spin2Int} &\\
	(\bar u_2 q^\alpha \sigma_{\alpha\mu_1}u_1)(\bar u_4 q^\beta \sigma_{\beta\mu_2}u_3) v_{1\mu_3\dots\mu_{2k}}\mathcal{I}_\triangleleft^{\mu_1\dots\mu_{2k}},\label{eq:Spin1Spin2Int} &\\
	S_{1\mu_{1}}S_{2\mu_{2}}v_{1\mu_{3}\dots\mu_{2k}}\mathcal{I}_{\triangleleft}^{\mu_{1}\dots\mu_{2k}}, &\\
	q\cdot S_{1}\,S_{2\mu_{1}}v_{1\mu_{2}\dots\mu_{2k+1}}\mathcal{I}_{\triangleleft}^{\mu_{1}\dots\mu_{2k+1}}=S_{1\mu_{1}}\,q\cdot S_{2}v_{1\mu_{2}\dots\mu_{2k+1}}\mathcal{I}_{\triangleleft}^{\mu_{1}\dots\mu_{2k+1}}. &
\end{align}
The leading terms for these integrals are
\begin{align}
	(\bar u_2 q^\alpha \sigma_{\alpha\mu_1}u_1) v_{1\mu_2\dots\mu_{2k}}\mathcal{I}_\triangleleft^{\mu_1\dots\mu_{2k}} &=
	(\bar u_2 q^\alpha v_2^\beta \sigma_{\alpha\beta} u_1) \frac{\left(\frac{1}{2}\right)_{k}}{4^{k}(1)_{k}} \omega(\omega^2-1)^{k-1} q^{2k}\mathcal{I}_\triangleleft , \\
	(\bar u_4 q^\alpha \sigma_{\alpha\mu_1}u_3) v_{1\mu_2\dots\mu_{2k}}\mathcal{I}_\triangleleft^{\mu_1\dots\mu_{2k}} &=
	-(\bar u_4 q^\alpha v_1^\beta \sigma_{\alpha\beta} u_3) \frac{\left(\frac{1}{2}\right)_k}{4^{k}(1)_k} (\omega^2-1)^{k-1} q^{2k}\mathcal{I}_\triangleleft , \\
	(\bar u_2 q^\alpha \sigma_{\alpha\mu_1}u_1)(\bar u_4 q^\beta \sigma_{\beta\mu_2}u_3) v_{1\mu_3\dots\mu_{2k}}&\mathcal{I}_\triangleleft^{\mu_1\dots\mu_{2k}}= \\
	(\bar u_2 q^\alpha \sigma_{\alpha\mu}u_1)  (\bar u_4 q^\beta \sigma_{\beta\nu} u_3)&\frac{\left(\frac{1}{2}\right)_{k}}{4^{k}(1)_{k}}q^{2k} \mathcal{I}_{\triangleleft}\notag
	\\
	\times\left[- \frac{1}{2k-1}\right.&\left.\eta^{\mu\nu} (\omega^2-1)^{k-1}-\frac{2(k-1)}{(2k-1)} v_{1}^{\nu}v_{2}^{\mu}\omega (\omega^2-1)^{k-2} \right],
	\nonumber \\
	S_{1\mu_{1}}S_{2\mu_{2}} v_{1\mu_3\dots\mu_{2k}}\mathcal{I}_\triangleleft^{\mu_1\dots\mu_{2k}}
	&=S_{1\mu_{1}}S_{2\nu}\frac{\left(\frac{1}{2}\right)_{k}}{4^{k}(1)_{k}}q^{2k} \mathcal{I}_{\triangleleft}
	\\
	\times\left[\frac{2k+1}{2k-1}\right.&\left.q^{\mu}q^{\nu}(\omega^{2}-1)^{k-1}- \frac{1}{2k-1}q^{2}\eta^{\mu\nu} (\omega^2-1)^{k-1}\right.\notag \\
	&\qquad\left.-\frac{2(k-1)}{(2k-1)} q^{2}v_{1}^{\nu}v_{2}^{\mu}\omega (\omega^2-1)^{k-2} \right],
	\nonumber\\
	q\cdot S_{1}\,S_{2\mu_{1}}v_{1\mu_{2}\dots\mu_{2k+1}}\mathcal{I}_{\triangleleft}^{\mu_{1}\dots\mu_{2k+1}}&=-q\cdot S_{1}\,q\cdot S_{2}\frac{\left(\frac{1}{2}\right)_{k}}{2^{2k+1}(1)_{k}}(\omega^{2}-1)^{k}q^{2k}\mathcal{I}_{\triangleleft}.
\end{align}
where $k\geq1$.
The first three of these contractions rely on the same portion of the rank-$2k$
triangle integral as that of \cref{eq:SpinlessTriIntGenK};
namely, only the part of the integral whose tensor structure contains no transfer momenta.
As such, these integrals can be derived by a combinatoric analysis of \cref{eq:SpinlessTriIntGenK}.
We show now this combinatoric analysis.

The leading-in-$\hbar$ portion of the triangle integral proportional to tensor structures
containing only factors of the metric and the velocity can be inferred from \cref{eq:SpinlessTriIntGenK}.
To do this, we expand the binomial $(\omega^{2}-1)^{k}$ and "uncontract" the integral by noting that
$\omega=v_{1\mu}v_{2}^{\mu}$ and $v_{1\mu\nu}\eta^{\mu\nu}=1$:
\begin{align}
    v_{1\mu_1\dots\mu_{2k}}&\mathcal{I}_\triangleleft^{\mu_1\dots\mu_{2k}}=\frac{\left(\frac{1}{2}\right)_{k}}{4^{k}(1)_k}q^{2k} \mathcal{I}_\triangleleft\sum_{n=0}^{k}{k\choose n}(-1)^{k-n}v_{1\mu_{1}\dots\mu_{2k-2n}}\eta^{\mu_{1}\mu_{2},\dots,\mu_{2k-2n-1}\mu_{2k-2n}}v_{1\alpha_{1}\dots\alpha_{2n}}v_{2}^{\alpha_{1}\dots\alpha_{2n}}\notag \\
    &=v_{1\mu_{1}\dots\mu_{2k}}\left[\frac{\left(\frac{1}{2}\right)_{k}}{4^{k}(1)_k}q^{2k} \mathcal{I}_\triangleleft\sum_{n=0}^{k}\frac{1}{(2k)!}{k\choose n}(-1)^{k-n}\eta^{\{\mu_{1}\mu_{2},\dots,\mu_{2k-2n-1}\mu_{2k-2n}}v_{2}^{\mu_{2k-2n+1}\dots\mu_{2k}\}}\right],
\end{align}
where $\eta^{\mu_{1}\mu_{2},\dots,\mu_{2k-1}\mu_{2k}}\equiv\eta^{\mu_{1}\mu_{2}}\dots\eta^{\mu_{2k-1}\mu_{2k}}$.
We now identify the quantity in the square brackets with the uncontracted triangle integral.
Curly brackets denote symmetrization without normalization.
Note that this is not the actual value of the uncontracted triangle integral: even at the
order of $\hbar$ at which we are working, tensor structures with an even number of transfer-momentum
four-vectors are present.
For our purposes, however, these contributions will always vanish or become subleading
through the on-shell condition $v_{1}\cdot q\sim q^{2}/m$ when contracted with the spin
structures in \cref{eq:Spin1Int,eq:Spin2Int,eq:Spin1Spin2Int}.

We begin with the contraction in \cref{eq:Spin1Int}.
\Cref{eq:Spin1Ident} implies that the $n=0$ term vanishes, and that the only non-vanishing terms
have a $v_{2}^{\mu}$ contracted with the sigma matrix.
The remaining indices will all be contracted symmetrically with factors of $v_{1\mu}$,
so for a term with $2n$ factors of $v_{2}^{\mu}$ and $2k$ total symmetrized Lorentz indices
there will be $2n(2k-1)!$ non-vanishing and identical distributions of the Lorentz indices.
Then,
\begin{align}
    &(\bar u_2 q^\alpha \sigma_{\alpha\mu_1}u_1) v_{1\mu_2\dots\mu_{2k}}\mathcal{I}_\triangleleft^{\mu_1\dots\mu_{2k}}\notag \\
    &=(\bar u_2 q^\alpha v_{2}^{\mu_{2k}} \sigma_{\alpha\mu_{2k}}u_{2k}) \frac{\left(\frac{1}{2}\right)_{k}}{4^{k}(1)_k}q^{2k} \mathcal{I}_\triangleleft\sum_{n=1}^{k}\frac{2n(2k-1)!}{(2k)!}{k\choose n}(-1)^{k-n}\omega^{2n-1}\notag \\
    &=(\bar u_2 q^\alpha v_{2}^{\mu_{2k}} \sigma_{\alpha\mu_{2k}}u_{2k}) \frac{\left(\frac{1}{2}\right)_{k}}{4^{k}(1)_k}q^{2k} \mathcal{I}_\triangleleft\omega\sum_{n=0}^{k-1}{{k-1}\choose {n}}(-1)^{k-1-n}\omega^{2n-2}\notag \\
    &=(\bar u_2 q^\alpha v_{2}^{\mu_{2k}} \sigma_{\alpha\mu_{2k}}u_{2k}) \frac{\left(\frac{1}{2}\right)_{k}}{4^{k}(1)_k}\omega(\omega^{2}-1)^{k-1}q^{2k} \mathcal{I}_\triangleleft.
\end{align}
This is what we wanted to prove.

The argument for the contraction in \cref{eq:Spin2Int} is similar.
However, because of \cref{eq:Spin2Ident}, now only terms with at least one metric survive.
Thus the $n=k$ term vanishes, and a term with $k-n$ metrics and $2k$ total symmetrized Lorentz indices
contributes $2(k-n)(2k-1)!$ identical Lorentz index distributions.
Therefore,
\begin{align}\label{eq:Spin2Proof1}
    	&(\bar u_4 q^\alpha v_{1}^{\mu_{1}} \sigma_{\alpha\mu_1}u_3) v_{1\mu_2\dots\mu_{2k}}\mathcal{I}_\triangleleft^{\mu_1\dots\mu_{2k}}\notag \\
    	&=	(\bar u_4 q^\alpha v_{1}^{\mu_{2k}} \sigma_{\alpha\mu_{2k}}u_3) \frac{\left(\frac{1}{2}\right)_{k}}{4^{k}(1)_k}q^{2k} \mathcal{I}_\triangleleft\sum_{n=0}^{k-1}\frac{2(k-n)(2k-1)!}{(2k)!}{k\choose {k-n}}(-1)^{k-n}\omega^{2n}.
\end{align}
We have used the identity ${k\choose {n}}={k\choose {k-n}}$.
Inverting the order of the sum,
\begin{align}
    	&(\bar u_4 q^\alpha v_{1}^{\mu_{1}} \sigma_{\alpha\mu_1}u_3) v_{1\mu_2\dots\mu_{2k}}\mathcal{I}_\triangleleft^{\mu_1\dots\mu_{2k}}\notag \\
    	&=	(\bar u_4 q^\alpha v_{1}^{\mu_{2k}} \sigma_{\alpha\mu_{2k}}u_3) \frac{\left(\frac{1}{2}\right)_{k}}{4^{k}(1)_k}q^{2k} \mathcal{I}_\triangleleft\sum_{n=1}^{k}\frac{2n(2k-1)!}{(2k)!}{k\choose {n}}(-1)^{n}\left(\omega^{2}\right)^{k-n}\notag \\
    	&=	(\bar u_4 q^\alpha v_{1}^{\mu_{2k}} \sigma_{\alpha\mu_{2k}}u_3) \frac{\left(\frac{1}{2}\right)_{k}}{4^{k}(1)_k}q^{2k} \mathcal{I}_\triangleleft\sum_{n=0}^{k-1}{{k-1}\choose {n}}(-1)^{n+1}\left(\omega^{2}\right)^{k-1-n}\notag \\
    	&=	-(\bar u_4 q^\alpha v_{1}^{\mu_{2k}} \sigma_{\alpha\mu_{2k}}u_3) \frac{\left(\frac{1}{2}\right)_{k}}{4^{k}(1)_k}\left(\omega^{2}-1\right)^{k-1}q^{2k} \mathcal{I}_\triangleleft,
\end{align}
as claimed.

Finally, we move to the contraction of \cref{eq:Spin1Spin2Int}.
\Cref{eq:Spin2Ident} implies that the $n=k$ term in the sum vanishes, and that we must
always have one metric contracted with the spin of particle 2.
There are then clearly two types of contributions:
those where the second index of that metric contracts
with the spin of particle 1, and those where it
contracts with a factor of $v_{1}^{\mu}$:
\begin{align}
    &(\bar u_2 q^\alpha \sigma_{\alpha\mu_1}u_1)(\bar u_4 q^\beta \sigma_{\beta\mu_2}u_3) v_{1\mu_3\dots\mu_{2k}}\mathcal{I}_\triangleleft^{\mu_1\dots\mu_{2k}}\notag \\
    &=(\bar u_2 q^\alpha \sigma_{\alpha\mu_1}u_1)(\bar u_4 q^\beta \sigma_{\beta\mu_2}u_3) v_{1\mu_3\dots\mu_{2k}}\frac{\left(\frac{1}{2}\right)_{k}}{4^{k}(1)_k}q^{2k} \mathcal{I}_\triangleleft\notag \\
    &\quad\times\left[\eta^{\mu_{1}\mu_{2}}\sum_{n=0}^{k-1}\frac{N_{n}}{(2k)!}{k\choose n}(-1)^{k-n}\eta^{\{\mu_{3}\mu_{4},\dots,\mu_{2k-2n-1}\mu_{2k-2n}}v_{2}^{\mu_{2k-2n+1}\dots\mu_{2k}\}}\right.\notag \\
    &\qquad\left.+\eta^{\mu_{2}\mu_{3}}\sum_{n=0}^{k-1}\frac{P_{n}}{(2k)!}{k\choose n}(-1)^{k-n}\eta^{\{\mu_{1}\mu_{4},\dots,\mu_{2k-2n-1}\mu_{2k-2n}}v_{2}^{\mu_{2k-2n+1}\dots\mu_{2k}\}}\right].
\end{align}
Our task is now to determine the integers $N_{n}$ and $P_{n}$ 
-- which count the number of identical Lorentz index
distributions that do not vanish when contracted
with the spin structure -- and to resum the sums.
In the case of $N_{n}$, in a term with $k-n$ metrics, there 
are $k-n$ metrics that can be contracted with the spin of particle 2, two identical ways to distribute
the indices on this metric, and $(2k-2)!$ identical ways to
distribute the remaining Lorentz indices for each of these metric permutations.
Therefore, $N_{n}=2(k-n)(2k-2)!$.

Now, in the term corresponding to $P_{n}$, \cref{eq:Spin1Ident} implies that the $n=0$ term also vanishes,
and that one factor of $v_{2}^{\mu}$ must be contracted with
the spin of particle 1.
For a term with $2n$ factors of $v_{2}^{\mu}$, there are then
$2n$ identical ways to contract a velocity with the spin.
Moreover, such a term will have $k-n$ metrics that can be contracted
with the spin of particle 2, each of which has two
indices that can be contracted in this way.
Finally, as \cref{eq:Spin1Ident,eq:Spin2Ident} fix two indices,
there are $(2k-2)!$ identical permutations of the remaining
Lorentz indices.
Therefore, $P_{n}=(2n)[2(k-n)](2k-2)!$.

Putting these together,
\begin{align}
    &(\bar u_2 q^\alpha \sigma_{\alpha\mu_1}u_1)(\bar u_4 q^\beta \sigma_{\beta\mu_2}u_3) v_{1\mu_3\dots\mu_{2k}}\mathcal{I}_\triangleleft^{\mu_1\dots\mu_{2k}}\notag \\
    &=(\bar u_2 q^\alpha \sigma_{\alpha\mu_{1}}u_1)(\bar u_4 q^\beta \sigma_{\beta\mu_2}u_3) \frac{\left(\frac{1}{2}\right)_{k}}{4^{k}(1)_k}q^{2k} \mathcal{I}_\triangleleft\notag \\
    &\quad\times\left[\eta^{\mu_{1}\mu_{2}}\sum_{n=0}^{k-1}\frac{(k-n)}{k(2k-1)}{k\choose n}(-1)^{k-n}(\omega^{2})^{n}\right.\notag \\
    &\qquad\left.+\eta^{\mu_{2}\mu_{3}}v_{1\mu_{3}}v_{2}^{\mu_{1}}\omega\sum_{n=0}^{k-2}\frac{2(n+1)(k-1-n)}{k(2k-1)}{k\choose {n+1}}(-1)^{k-1-n}(\omega^{2})^{n}\right].
\end{align}
We recognize the first sum as the one in \cref{eq:Spin2Proof1} divided by $2k-1$.
Focusing on the second sum, we apply the recursive identity for the binomial coefficients, invert
the sum, multiply by $(k-1)/(k-1)$, and apply the recursive identity again to find
\begin{align}
    &(\bar u_2 q^\alpha \sigma_{\alpha\mu_1}u_1)(\bar u_4 q^\beta \sigma_{\beta\mu_2}u_3) v_{1\mu_3\dots\mu_{2k}}\mathcal{I}_\triangleleft^{\mu_1\dots\mu_{2k}}\notag \\
    &=(\bar u_2 q^\alpha \sigma_{\alpha\mu}u_1)(\bar u_4 q^\beta \sigma_{\beta\nu}u_3) \frac{\left(\frac{1}{2}\right)_{k}}{4^{k}(1)_k}q^{2k} \mathcal{I}_\triangleleft\notag \\
    &\quad\times\left[-\frac{\eta^{\mu\nu}}{(2k-1)}(\omega^{2}-1)^{k-1}+v_{1}^{\nu}v_{2}^{\mu}\omega\sum_{n=0}^{k-2}\frac{2(k-1)}{(2k-1)}{{k-2}\choose {n}}(-1)^{n+1}(\omega^{2})^{k-n-2}\right]\notag \\
    &=(\bar u_2 q^\alpha \sigma_{\alpha\mu}u_1)(\bar u_4 q^\beta \sigma_{\beta\nu}u_3) \frac{\left(\frac{1}{2}\right)_{k}}{4^{k}(1)_k}q^{2k} \mathcal{I}_\triangleleft\notag \\
    &\qquad\times\left[-\frac{\eta^{\mu\nu}}{(2k-1)}(\omega^{2}-1)^{k-1}-\frac{2(k-1)}{(2k-1)}v_{1}^{\nu}v_{2}^{\mu}\omega(\omega^{2}-1)^{k-2}\right].
\end{align}
This completes the proof.

All integrals in this section were checked explicitly up to rank $2k=10$.

\section{Integrals for classical impulses and the potential}\label{app:fourierInt}

We give here details about the integrals needed to Fourier transform the potential
in \cref{sec:Hamiltonian} to position space,
as well as the integrals used to derive the linear and angular impulses in
\cref{sec:impulse,sec:eikonal}.

\subsection{Fourier transforms}

To convert the momentum space tidal potential in \cref{sec:Hamiltonian} to position space,
we need knowledge of three-dimensional Fourier integrals up to rank 2:
\begin{align}
    J^{i_{1}\dots i_{k}}_{j}&=-\frac{1}{\sqrt{2}}\int\hat{d}^{3}\bm{q}\ e^{-i\bm{q}\cdot \bm{r}}\left(\frac{\bm{q}^{2}}{2}\right)^{j+3/2}\bm{q}^{i_{1}\dots i_{k}},\quad k\leq 2.
\end{align}
First, for $k=0$,
\begin{align}
    J_{j}&=2\left(\frac{2}{r^{2}}\right)^{j+3}\frac{\Gamma\left(j+3\right)}{(4\pi)^{3/2}\Gamma(-j-3/2)},
\end{align}
where $r\equiv|\bm{r}|$.
In the rank 1 case,
\begin{align}
    J_{j}^{i}&=i\frac{\partial}{\partial \bm{r}^{i}}J_{j}=-2i\bm{r}^{i}\left(\frac{2}{r^{2}}\right)^{j+4}\frac{\Gamma\left(j+4\right)}{(4\pi)^{3/2}\Gamma(-j-3/2)}.
\end{align}
Similarly, for rank 2,
\begin{align}
    J^{ik}_{j}&=i\frac{\partial}{\partial \bm{r}^{k}}J^{i}_{j}=2\left(\frac{2}{r^{2}}\right)^{j+4}\left[\delta^{ik}-2(j+4)\frac{\bm{r}^{i}\bm{r}^{k}}{r^{2}}\right]\frac{\Gamma\left(j+4\right)}{(4\pi)^{3/2}\Gamma(-j-3/2)}.
\end{align}

\subsection{Linear and angular impulse}

The calculation of the classical impulses in \cref{sec:impulse,sec:eikonal} needed the evaluation of integrals of the form
\begin{align}\label{eq:ClassInt}
	I_{j+2}^{\mu_1\dots\mu_k}(b) =
	\int \hat d^4 q \hat \delta(2p_1\cdot q)\hat\delta(2p_2\cdot q)
	e^{-ib\cdot q} q^{\mu_1\dots\mu_k} S  \left(-\frac{q^2}{2}\right)^{j+2} ,
\end{align}
where $\hat d^n q = d^n q /(2\pi)^n$ and $\hat\delta(x)=2\pi \delta(x)$.
These integrals are similar to integrals evaluated in refs.~\cite{Kosower:2018adc,Maybee:2019jus},
and we will use the same steps here.
We go to the rest frame of particle 1, where $v_1 = (1,0,0,0)$ and $v_2 = (\omega,0,0,\omega\beta)$, with $\beta$ satisfying $\omega^2(1-\beta^2)=1$.
After evaluating the delta functions, we find that
\begin{align}
		I_{j+2}^{\mu_1\dots\mu_k}(\bm{b}) =& \frac{1}{\mathcal{N}}
		\int d^2 q e^{i\bm{b}\cdot \bm{q}}
		q^{\mu_1\dots\mu_k}\frac{1}{|\bm{q}|} \left( \frac{\bm{q}^2}{2}\right)^{j+2},
			\quad \text{where} \quad
		\mathcal{N} \equiv 16 m_1 m_2 \sqrt{\omega^2-1} .
\end{align}
We will omit the argument of the integral from now on.
We parametrize $\bm{q}$ to be in the 2-dimensional plane perpendicular to $v_1$ and $v_2$,
$ q^\mu = (0,\chi \cos\theta,\chi\sin\theta,0)$, such that the rank-0 integral becomes
\begin{align}
	I_{j+2} =& \frac{1}{\mathcal{N}} \int d^2 q e^{i \bm{b}\cdot \bm{q}}
	\frac{1}{|\bm{q}|} \left(\frac{\bm{q}^2}{2}\right)^{j+2}
	\nonumber \\
	=& \frac{1}{\mathcal{N}} \int_0^\infty d\chi \chi \int_{-\pi}^{\pi}d\theta e^{i |\bm{b}|\chi \cos\theta}
	\frac{1}{\chi} \left(\frac{\chi^2}{2}\right)^{j+2}
	\nonumber \\
	=& \frac{2\pi}{\mathcal{N}} \int_0^\infty d\chi J_0(|\bm{b}|\chi) 
	\left(\frac{\chi^2}{2}\right)^{j+2}
	\nonumber \\
	=& \frac{\pi}{8m_1 m_2 \sqrt{\omega^2-1}} \frac{1}{|\bm{b}|} \left(-\frac{2}{b^2}\right)^{j+2} \frac{\Gamma[5/2+j]}{\Gamma[-3/2-j]} ,
\end{align} 
where in the last step we restored Lorentz invariance.
The rank-1 integral is also straight-forward to evaluate;
\begin{align}
	\label{eq:Irank1result}
	I_{j+2}^{\mu} &=
	\frac{1}{\mathcal{N}}
	\int_0^\infty d\chi \chi \int_{-\pi}^{\pi} d\theta
	e^{i|\bm{b}|\chi\cos\theta} q^{\mu} \frac{1}{\chi}  \left(\frac{\chi^2}{2}\right)^{j+2}  \nonumber \\
		      &=\frac{2\pi i}{\mathcal{N}}
	\int_0^\infty d\chi \chi
	J_1(|\bm{b}|\chi) \bm{\hat b}^\mu   \left(\frac{\chi^2}{2}\right)^{j+2}  \nonumber \\
	&=\frac{\pi i}{8 m_1 m_2 \sqrt{\omega^2-1} }
	\frac{b^\mu}{|\bm{b}|}  \left(-\frac{2}{b^2}\right)^{j+3}
	 \frac{\Gamma[7/2+j]}{\Gamma[-3/2-j]} .
\end{align}
Next, we have the rank-2 integral, which must take the form
\begin{align}
	\label{eq:Irank2}
	I_{j+2}^{\mu\nu} = \alpha_2 b^\mu b^\mu + \beta_2 \Pi^{\mu\nu},
\end{align}
where
\begin{align}
	\label{eq:PiProjector}
	\Pi^\mu_{\,\,\nu} = \delta^\mu_{\,\,\nu}
	+ \frac{1}{\omega^2-1}\left(
	v_1^\mu (v_{1\nu}-\omega v_{2\nu}) + v_2^\mu (v_{2\nu}-\omega v_{1\nu})
	\right)
\end{align}
is the projector onto the 2-dimensional plane perpendicular to $v_1$ and $v_2$ \cite{Maybee:2019jus}.
By taking the trace of \cref{eq:Irank2}, we have that
\begin{align}
	\label{eq:traceIrank2}
	\alpha_2 b^2 + 2\beta_2  &=
	-\frac{1}{\mathcal{N}} \int_0^{\infty}
	d\chi \chi \int_{-\pi}^{\pi} d\theta
	e^{i|\bm{b}|\bm{q}\cos\theta}  \chi \left(
		\frac{\chi^2}{2}
	\right)^{j+2}
	\nonumber \\
				 &= -\frac{4\pi}{\mathcal{N}} \int_0^{\infty}
	d \chi 
	J_0(|\bm{b}|\chi)
	\left(
		\frac{\chi^2}{2}
	\right)^{j+3}
	\nonumber \\
				 &= -\frac{4\pi}{\mathcal{N}}
	\frac{1}{|\bm{b}|}
	\left(-\frac{2}{b^2}\right)^{j+3}
	\frac{\Gamma[7/2+j]}{\Gamma[-5/2-j]}.
\end{align}
If we contract with $b_{\mu}b_{\nu}$ instead, we find that
\begin{align}
	\label{eq:bbIrank2}
	\alpha_2 b^4 
	+ \beta_2 b^2
	&=
	-\frac{b^2}{\mathcal{N}}
	\int_0^\infty d\chi \chi^2 \int_{-\pi}^{\pi} d\theta
	e^{i|\bm{b}|\chi\cos\theta} (\cos\theta)^2   \left(\frac{\chi^2}{2}\right)^{j+2}  \nonumber \\
	&=
	-\frac{4\pi b^2}{\mathcal{N}}
	\int_0^\infty d\chi
	\left(J_1(|\bm{b}|\chi)\frac{1}{|\bm{b}|\chi}
		- J_2(|\bm{b}|\chi)
	\right)
		\left(\frac{\chi^2}{2}\right)^{j+3}  \nonumber \\
	&=
	\frac{4\pi b^2}{\mathcal{N}}
	\frac{1}{|\bm{b}|}
	 \left(-\frac{2}{b^2}\right)^{j+3}
	 \frac{(j+3)\Gamma[7/2+j]}{\Gamma[-3/2-j]} .
\end{align}
Putting this together with \cref{eq:traceIrank2}, we end up with
\begin{align}
	\label{eq:Irank2result}
	I_{j+2}^{\mu\nu} =
	\left(
		(7+2j)\frac{b^\mu b^\nu }{b^2}
	- \Pi^{\mu\nu}
	\right)
	\frac{\pi}{8 m_1 m_2 \sqrt{\omega^2-1}}
	\frac{1}{|\bm{b}|}
	 \left(-\frac{2}{b^2}\right)^{j+3}
	 \frac{\Gamma[7/2+j]}{\Gamma[-3/2-j]}.
\end{align}
Lastly, the rank-3 integral is
\begin{align}
	I^{\mu\nu\rho}_{j+2} = \alpha_3 b^\mu b^\nu b^\rho + \beta_3 b^{(\mu}\Pi^{\nu\rho)}
\end{align}
(with normalization $1/3!$ for the second term).
Contracting with $b_\mu\eta_{\nu\rho}$, we have that
\begin{align}
	\alpha_3 b^4 + \beta_3 \frac{4}{3} b^2 &=
	\frac{1}{\mathcal{N}}
	\int_0^\infty d\chi \chi \int_{-\pi}^{\pi} d\theta
	e^{i|\bm{b}|\chi\cos\theta} |\bm{b}| \chi^2 \cos\theta \left(\frac{\chi^2}{2}\right)^{j+2}  \nonumber \\ &=
			\frac{4\pi i|\bm{b}|}{\mathcal{N}}
		\int_0^\infty d\chi J_1(|\bm{b}|\chi ) \chi  \left(\frac{\chi^2}{2}\right)^{j+3}  \nonumber \\ &=
		\frac{8\pi i}{\mathcal{N}}
	 \frac{1}{|\bm{b}|} \left(\frac{2}{\bm{b}^2}\right)^{j+3}  \frac{\Gamma[9/2+j]}{\Gamma[-5/2-j]} .
\end{align}
When we contract with $b_\mu b_\nu b_\rho$, we find that
\begin{align}
	\alpha_3 b^6
	+ \beta b^4
	&=
	-\frac{1}{\mathcal{N}}
	\int_0^\infty d\chi \chi \int_{-\pi}^{\pi} d\theta
	e^{i|\bm{b}|\chi\cos\theta} (|\bm{b}|\chi\cos\theta)^3 \frac{1}{\chi}  \left(\frac{\chi^2}{2}\right)^{j+2}  \nonumber \\ &=
	-\frac{2\pi i}{\mathcal{N}}
	|\bm{b}|^3
	\int_0^\infty d\chi
	\left(
		J_2(|\bm{b}|\chi) \frac{3}{|\bm{b}|\chi}
		- J_3(|\bm{b}\chi)
	\right)
	\chi^3
	\left(\frac{\chi^2}{2}\right)^{j+2}   
	\nonumber \\ 
																 &=
																 -\frac{8\pi i b^2}{\mathcal{N}} \frac{1}{|\bm{b}|}
																 \left(-\frac{2}{b^2}\right)^{j+3}
																 \frac{(j+3)\Gamma[9/2+j]}{\Gamma[-3/2-j]} .
\end{align}
The integral becomes
\begin{align}
	\label{eq:Irank3result}
	I^{\mu\nu\rho}_{j+2} =
	\left(
	(9/2+j)\frac{b^\mu b^\nu b^\rho}{b^2}
	-
\frac{3}{2}b^{(\mu}\Pi^{\nu\rho)}\right)
	\frac{i\pi}{4 m_1 m_2 \sqrt{\omega^2-1}}
	\frac{1}{|\bm{b}|}
	\left(-\frac{2}{b^2}\right)^{j+4}  \frac{\Gamma[9/2+j]}{\Gamma[-3/2-j]} .
\end{align}

We also need these integrals restricted to spatial indices.
We can write the spatial portions of the integrals in as
\begin{align}
    I^{i_{1}\dots i_{n}}_{j+2}=(-i)^{n}\frac{\partial^{n}}{\partial b^{i_{1}}\dots\partial b^{i_{n}}}I_{j+2}.
\end{align}
Computing derivatives of the rank 0 integral gives
\begin{align}
    I^{m}_{j+2}&=\frac{\pi i}{8 m_1 m_2 \sqrt{\omega^2-1} }
	\frac{\bm{b}^m}{|\bm{b}|}  \left(\frac{2}{\bm{b}^2}\right)^{j+3}
	 \frac{\Gamma[7/2+j]}{\Gamma[-3/2-j]}, \\
	 I^{mn}_{j+2}&=\left(
		\Pi^{mn}-(7+2j)\frac{\bm{b}^m \bm{b}^n }{\bm{b}^2}
	\right)
	\frac{\pi}{8 m_1 m_2 \sqrt{\omega^2-1}}
	\frac{1}{|\bm{b}|}
	 \left(\frac{2}{\bm{b}^2}\right)^{j+3}
	 \frac{\Gamma[7/2+j]}{\Gamma[-3/2-j]}, \\
	 I^{mnl}_{j+2}&=\left(\frac{3}{2}\bm{b}^{(m}\Pi^{nl)}-
	(9/2+j)\frac{\bm{b}^m \bm{b}^n \bm{b}^l}{\bm{b}^2}
	\right)
	\frac{i\pi}{4 m_1 m_2 \sqrt{\omega^2-1}}
	\frac{1}{|\bm{b}|}
	\left(\frac{2}{\bm{b}^2}\right)^{j+4}  \frac{\Gamma[9/2+j]}{\Gamma[-3/2-j]}.
\end{align}
Moreover, note that the projector to the plane perpendicular to the velocities must be modified
when restricted to purely spatial components:
\begin{align}
    \Pi^{mn}&=\delta^{mn}- \frac{1}{\omega^2-1}\left[
	\bm{v}_1^m (\bm{v}_{1}^{n}-\omega \bm{v}_{2}^{n}) + \bm{v}_2^m (\bm{v}_{2}^{n}-\omega \bm{v}_{1}^{n})
	\right].
\end{align}

\subsection{Eikonal operator}

The eikonal phase was written in \cref{sec:eikonal}
as the action of the operator $\hat{\mathcal{K}}_j$ on the integral $I_{j+2}$.
We list here for reference the action of each term in this operator on this integral:
\begin{align}
\label{eq:EikonalOpTerms}
\hat{\mathcal{O}}^{(0)} I_{j+2} &= I_{j+2},\\
\hat{\mathcal{O}}^{(1,a)} I_{j+2} &=- i\,( \bm{S}_a \times \bm{p})_m I^m_{j+2},\\
\hat{\mathcal{O}}^{(2,1)} I_{j+2} &= \bm{S}_{1m} \bm{S}_{2n} I^{mn}_{j+2},\\
\hat{\mathcal{O}}^{(2,2)} I_{j+2} &= (\bm{S}_1 \cdot \bm{S}_2) \delta_{mn} I^{mn}_{j+2} =2 (\bm{S}_1 \cdot \bm{S}_2) I_{j+3},\\
\hat{\mathcal{O}}^{(2,3)} I_{j+2} &= (\bm{p} \cdot \bm{S}_1)(\bm{p} \cdot \bm{S}_2) \delta_{mn} I^{mn}_{j+2} = 2 (\bm{p} \cdot \bm{S}_1)(\bm{p} \cdot \bm{S}_2)  I_{j+3},
\end{align}
where $m,n$ are spatial indices.

\bibliographystyle{JHEP}
\bibliography{SpinningTidalEffects}

\end{document}